\documentclass[a4paper,10pt]{article}
\usepackage{amsmath,amssymb,amsthm}
\usepackage{dsfont}
\theoremstyle{definition}

\theoremstyle{definition}

\theoremstyle{definition}

\theoremstyle{remark}

\begin{document}
\title{Berezin integration over anticommuting variables and cyclic cohomology}
\author{G. Grensing and M. Nitschmann\\ Institut f\H ur Theoretische Physik und Astrophysik \\ 
Universit\H at Kiel \\D-24118 Kiel, 
Germany}

\date{January 2004}

\maketitle

\begin{abstract}
Berezin integration over fermionic degrees of freedom as a standard tool of quantum field theory is analysed from the viewpoint of noncommutative geometry. It is shown that among the variety of contradictory integration prescriptions existing in the current literature, there is only one unique minimal set of consistent rules, which is compatible with Connes' normalized cyclic cohomology of the Gra{\ss}mann algebra. 
\end{abstract}
\thispagestyle{empty}
\newpage
\setcounter{page}{1}
\section*{Introduction}

That branch of mathematics in which a $\mathds{Z}/2$-grading plays dominant role is, somewhat  
emphatically, also called `supermathematics'. It is of fundamental importance in physics since it underlies the treatment of fermionic particles, and as such it is part of the standard repertoire in quantum field theory (see, e.g. \cite{Itzy80,Ryde85}). Of course, these tools get involved also in supersymmetric theories \cite{Nieu81,Wess83}, but for its justification there is no need to take recourse to supersymmetry since already for conventional theories containing fermions one must make use of what is known as the differential and integral calculus on a 
Gra{\ss}mann algebra. After preparatory work of Schwinger \cite{Schw53,Schw70}, 
Matthews $\&$ Salam \cite{Matt55} 
and others this calculus was created by the Russian mathematician Berezin \cite{Bere66} in the year 
1966. Since then this theory has undergone revision, also by Berezin himself (see \cite{Bere79,Bere87}),  and nowadays a variety of conflicting rules is given in the literature, 
e.g., some authors use anticommuting differentials for the Grassmann variables, whereas others use commuting ones, and the like. Hence, as regards the present status of the art concerning the Berezin calculus, the situation is controversial, to say the least.

In all, it seems that a more fundamental understanding of the Berezin rules is desirable. As we believe, with the advent of noncommutative geometry \`a la Connes \cite{Conn94} new avenues have been opened to reconsider these problems. In this context, a universal differential calculus over a general 
noncommutative algebra is supplied for, where the algebra is a substitute for the commutative algebra of functions over a manifold in the classical situation; furthermore, the notion of Connes' characters provides for abstract integral calculi over such a `virtual manifold'. Whence, this whole machinery can serve as an ideal vehicle to gain deeper insight into the origin of Berezin integration, which is the main motivation for the present work. 

The Gra{\ss}mann algebra is certainly noncommutative, but in a mild form since it is graded commutative. Thus, the extension of the universal differential algebra to the $\mathds{Z}/2$-graded situation is needed; this is available since the year 1988 with the book of Kastler \cite{Kast88}. His treatment essentially relies on the Karoubi \cite{Karo82} approach, which allows for a drastic simplification of the formalism. It  also delivers a rather natural set of sign factors that get involved in the definition of the Hochschild boundary operator and the cyclicity operator. But the latter differ from those obtained by means of the sign rules ascribed to Koszul, Milnor or Quillen, depending on ones preferences. In the subsequent work of Kastler and collaborators \cite{Coqu90,Coqu91,Coqu95}, the original sign factors were forgotten in favour of the customary Koszul-Milnor-Quillen signs. Though one would expect the choice of a sign to be a matter of convention, we will show that only the sign factors obtained by means of the Karoubi approach are compatible with the rules governing the Connes' characters of the Gra{\ss}mann algebra. 

Further progress was made in 1995 by Coquereaux $\&$ Ragoucy 
\cite{Coqu95} with their work on the cohomology of the Gra{\ss}mann algebra. They introduced the 
Gra{\ss}mann analogue of de Rham currents, which in the classical situation over a compact manifold were shown by Connes 
\cite{Conn85} to be in canonical one-to-one correspondence with skewsymmetrized Hochschild cohomology. One expects something similar to be valid in the Gra{\ss}mann case, and the above authors were partially successful in obtaining an analogous result. We shall sharpen their arguments and prove that Gra{\ss}mann  currents compute the normalized cyclic Hochschild cohomology of the Gra{\ss}mann algebra.

In defining a Gra{\ss}mann current, use is made of the Berezin rules, which thus have to be given beforehand. They are of completely algebraic origin and are specified by a linear map  $J:G^n\to\mathds{R}$, where $G^n$ denotes the  Gra{\ss}mann algebra with $n$ generators $\xi^i$, which we  require to be translational invariant. The explicit form of this condition will be specified later; 
it selects one unique map amongst this multitude of possibilities, the one that singles out the top component of an algebra element $f(\xi)\in G^n$, i.e. $J(f)=f_{1\dots n}$ where the right hand side denotes the coefficient of the term of highest degree. It has become customary to write this unique element in the suggestive form of an integral, namely
$$J(f)=\int d^{\,n}\xi\,f(\xi).$$
In this version, however, the meaning of the symbol $d^{\,n}\xi$ is left open, and up to the present day it awaits a proper definition. This is the main source for the many conflicting Berezin integration rules existing in the literature. It is the purpose of the present paper to justify the suggestive symbolic notation, and for this the abstract apparatus of noncommutative geometry is needed, in particular, the universal differential calculus and its associated cyclic cohomology (see, e.g., \cite{Land97,Grac01}). 

We understand the Berezin rules as a purely algebraic recipe, and only use it in this sense in the course of the development. This conception permits to ask the question whether Connes' characters have something to tell us about the deeper origin of the Berezin rules; the answer will be given at the very end of the present paper, where it is shown that there is one unique normalized cyclic cocycle that entails those Berezin rules, we except as valid. Furthermore, it provides for a proper definition of the 
`volume element' $d^{\,n}\xi$ that enters the formal Berezin integral. 

The paper is organized as follows: In the first Section, the universal differential graded algebra for a $\mathds{Z}/2$-graded algebra is constructed along with the definition of its normalized cyclic cohomology;  afterwards, the necessary modifications for a graded commutative algebra are given. The second Section supplies for the necessary basic facts about Gra{\ss}mann algebras; in particular, we give a proof that the request for translational invariance determines the Berezin rules uniquely. Furthermore, Fourier transformation is discussed and used to endow the Grassmann algebra with two hermitean inner products, both of which are nondegenerate. One of these is positive definite and provides for a Hilbert space structure, whereas the second one is indefinite; we shall construct a unitary grading operator, which relates these two inner products. They give rise to two operators, that are dual to the exterior differential operator, and these in turn are instrumental for a proof of the Poincar\'e lemma and the Hodge decomposition theorem. We then turn to Connes' characters over the Gra{\ss}mann algebra in the third Section and demonstrate that compatibility requires the use of the sign rules deriving from the Karoubi approach, whereas the Koszul-Milnor-Quillen conventions lead to contradictions. The indefinite inner product mentioned above gives rise to a nondegenerate pairing between the space of $p$-forms and $p$-currents; it is used in the fourth  Section to compute the normalized cyclic cohomology of the 
Gra{\ss}mann algebra by means of currents, where we use results derived earlier by Kassel \cite{Kass86}. These tools are applied in the fifth Section to single out among the possible Connes' characters one single element, which may legitimately be called a `volume character'. On the one hand, it embodies the  Berezin rules in an encoded form, and on the other hand it ultimately provides for a rigorous justification for the rewriting of the formal Berezin integral as an integral. We end in the last Section with some concluding remarks, being devoted to the impact of the above results on supersymmetric theories.   

\section{Connes' cyclic cohomology for $\mathds{Z}/2$-graded algebras}

We give a short introduction to the differential graded algebra of a $\mathds{Z}/2$-graded unital algebra $A$ over a commutative ring $k$, and the associated Connes' characters; we mainly follow the treatment given by Kastler \cite{Kast88,Kast90}, being based on the Karoubi approach \cite{Karo82}. This will prove to be essential in avoiding the nightmare to invent the correct Koszul-Milnor-Quillen signs, at least in part.

\subsection{The universal differential calculus} 

So let $A$ be an associative algebra with unit $e$ over a commutative ring $k$; it is also assumed to be  $\mathds{Z}/2$-graded, with $\varepsilon$ denoting the grading automorphism. The graded differential algebra, written $\Omega^{\displaystyle\cdot}A$, is constructed as follows through generators and relations.
We begin by defining the space of $1$-forms $\Omega_{}^1A$ on associating with every element $a\in A$ a symbol $da$; they generate the left $A$-module $\Omega_{}^1A$, and so we have a $k$-linear map $d:A\to \Omega_{}^1A$ which is required to obey the graded Leibniz rule
\begin{equation}
d(aa')=(da)a'+(-1)^{|d|\,|a|}a(da')\qquad\qquad\qquad:\,a,a'\in A 
\end{equation}
where $|a|$ denotes the parity. In this way, $\Omega_{}^1A$ is equipped also with a right $A$-structure. Since $|ea|=|e|+|a|$ implies $|e|=0$, the map $d$ annihilates the unit. The $A$-bimodule  $\Omega^pA$ of 
$p$-forms is defined as the tensor product $\Omega^pA=\overset{p}{\otimes}^{}_A\Omega_1A$, and its general element $\omega_p$ can thus be written as a finite linear combination of monomials $a_0da_1\cdots da_p$ with the tensor product sign being omitted; the structure of $\Omega^pA$ as a right $A$-module is determined by the easily verifiable identity
\begin{equation}
(-1)^pa_0da_1\cdots da_p\,a_{p+1}=
\end{equation}
\begin{equation}\nonumber
(-1)^{\sum\limits_{j=1}^p|a_j|}a_0a_1da_2\cdots da_{p+1}
+\sum_{i=1}^p(-1)^{i+\sum\limits_{j=i+1}^p|a_j|}a_0da_1\cdots d(a_ia_{i+1})\cdots da_{p+1}.
\end{equation}
Furthermore, the operator $d$ of exterior differentiation is defined by $d(a_0da_1\cdots da_p)=da_0da_1\cdots da_p$, which gives
$d(da_1\cdots da_p)=0$ and so we have 
$d^{\,2}=0$ on $\Omega^{\displaystyle\cdot}A=\underset{p}{\oplus}\Omega^pA$. Next we need  
$$d(a_0da_1\cdots da_p\cdot a'_0da'_1\cdots da'_p)=d(a_0da_1\cdots da_p)a'_0da'_1\cdots da'_p+
(-1)^{(p+|a_0|+\cdots+|a_p|)}a_0da_1\cdots da_p\,d(a'_0da'_1\cdots da'_p)$$
and thus the parity in the $\mathds{Z}/2$-graded case must be identified as
\begin{equation}
|a_0da_1\cdots da_p|=p+\sum_{i=0}^p|a_i|\qquad\qquad\qquad:\,|d|=1.
\end{equation}
Hence, the product rule is 
\begin{equation}
d(\omega\,\omega')=d\omega\,\omega'+(-1)^{|\omega|}\omega\,d\omega'
\end{equation}
which implies $d^{\,2}(\omega\,\omega')=0$, and so the assignment of the parity $|d|=1$ to the exterior derivation $d$ is indeed correct. Finally, the universality property of  $\Omega^{\displaystyle\cdot}A$ is proven in much the same way as in the trivially graded case (see, e.g. \cite{Land97}). 

We may then address the problem of defining a boundary operator on the universal differential graded algebra.
This is done, instead of starting from scratch, on taking the Karoubi form (\cite{Karo82}, see also 
\cite{Brod98}) of the ungraded case as guiding principle. Accordingly, the definition simply is
\begin{equation}
b(\omega\,da)=(-1)^{|\omega|}[\omega,a]\qquad\qquad\qquad\qquad b(a)=0
\end{equation}
where on the right hand side the graded commutator is understood; with the notation $a_0da_1\cdots da_{p}=(a_0,a_1,\ldots a_{p})$ the explicit form reads
\begin{equation}\label{boundopvarphi}
b(a_0,a_1,\ldots,a_{p})=
\end{equation}
\begin{equation}\nonumber
\sum_{i=0}^{p-1}(-1)^{i+\sum\limits_{j=0}^i|a_j|}(a_0,\ldots,a_ia_{i+1},\ldots,a_{p})
-(-1)^{(1+|a_{p}|)((p-1)+\sum\limits_{j=0}^{p-1}|a_j|)}(a_{p}\,a_0,a_1,\ldots,a_{p-1}).
\end{equation}
It reduces to the correct boundary operator in the trivially graded case and as well obeys $b^{\,2}=0$  in the present situation.

Let us also introduce a mutilated version of the boundary operator, denoted $b'$, in which only the crossover term is absent, viz.
\begin{equation}
b'(\omega\,da)=(-1)^{|\omega|}\omega\,a
\end{equation}
and $b'(a)=0$. It also obeys $b^{\,\prime 2}=0$, and its anticommutator with $d$ is 
\begin{equation}\label{anticdbprimeequunity}
b^{\,\prime}\,d+d\,b^{\,\prime}=1.
\end{equation}
Hence $b'$ is a contracting homotopy so that the complex $(\Omega^{\displaystyle\cdot}A,d)$ is acyclic.

In order to find the generalization of the cyclicity operator, we pass to the dual situation. A cochain $\varphi$ is a $k$-linear form on $\Omega^{\displaystyle\cdot}A$; 
it is called closed if $\varphi\circ d=0$, i.e. vanishes on exact forms, and graded if $\varphi\circ\text{ad}(\omega)=0$ for all  $\omega\in\Omega^{\displaystyle\cdot}A$, i.e. vanishes on graded commutators. So let $\varphi$ denote a both closed and graded trace; from $\varphi([\omega,da])=0$ we infer
\begin{align*}
\varphi(\omega\,da)&=(-1)^{|\omega|(1+|a|)}\varphi(da\,\omega)\\
&=(-1)^{|\omega|(1+|a|)}(\varphi(d(a\,\omega))-(-1)^{|a|}\varphi(a\,d\omega))\\
&=(-1)^{(1+|\omega|)(1+|a|)}\varphi(a\,d\omega)\\
&=\varphi(\lambda(a\,d\omega)))
\end{align*}
where the cyclicity operator acting on $\Omega^{\displaystyle\cdot}A$ is read off to be
\begin{equation}\label{cyclopvarphi}
\lambda(\omega\,da)=(-1)^{(1+|\omega|)(1+|a|)}a\,d\omega
\end{equation}
and we thus define it on cochains by the rule 
\begin{equation}\label{deflambdagng}
\lambda\,\varphi=\varphi\circ\lambda.
\end{equation}
Analogously, for the boundary operator, the definition is 
\begin{equation}\label{defboundopgng}
b\,\varphi=\varphi\circ\,b.
\end{equation}
In particular, if $\varphi$ is a graded trace, we thus have $b\,\varphi=0$. Finally, the definition of a character $\tau(a_0,a_1,\ldots a_{n})$ of an $n$-dimensional cycle $(\Omega^{(n)}A,d,\int)$ is the same as in the trivially graded case; hence, the cyclicity property for a character reads
\begin{equation}\label{cyclopkdefgncg}
\tau(a_0,\ldots,a_n)=(-1)^{(1+|a_n|)(n+\sum\limits_{i=0}^{n-1}|a_i|)}\tau(a_n,a_0,\ldots,a_{n-1})
\end{equation}
and, according to the previous remark, the boundary of $\tau$ vanishes. Furthermore, a slight variant of the proof in the trivially graded case (see, e.g. \cite{Grac01}) shows that the correspondence between characters and normalized cyclic Hochschild cocycles is one-to-one; we recall, a $p$-cochain $\varphi$ is called cyclic if $\lambda\,\varphi=\varphi$, and normalized if $\varphi(a^{}_0,\ldots,a_p)=0$ in case that $a_i=e$ for some $i\in\{0,\ldots,p\}$. It is in this way that Connes' cyclic cohomology can naturally be extended to the $\mathds{Z}/2$-graded case.

As for our notational conventions, the normalized cyclic cohomology is denoted by $HC^{\displaystyle\cdot}(A)$, whereas $H^{\displaystyle\cdot}_{\lambda}(A)$ signifies the cyclic cohomology. 

\subsection{The Koszul-Milnor-Quillen sign conventions}

The definition of the boundary operator and the cyclicity operator in the $\mathds{Z}/2$-graded case, as given throughout in the literature (see \cite{Conn85}), differs from the one we have obtained above. In order to relate these two versions (cf. also 
\cite{Kast88}), the heuristic motive consists in distributing $p$ factors of $d$ in front of the product $a_0a_1\cdots a_p$ such that one ends up 
with the $p$-form $a_0da_1\cdots da_p$; this gives the sign factor
\begin{equation}
\epsilon_p(a_0,a_1,\ldots,a_p)=(-1)^{\sum\limits_{i=0}^{p-1}(p-i)|a_i|}
\end{equation}
(where the dependence on the parities of the $a$s is suppressed below) and we thus define
\begin{equation}
\phi(a_0da_1\cdots da_p)=\epsilon_p\,\varphi(a_0da_1\cdots da_p).
\end{equation}
The new cochain $\phi$ inherits the cyclicity operator
$\lambda\phi(a_0da_1\cdots da_p)=\epsilon_p\,\lambda\varphi(a_0da_1\cdots da_p)$, 
denoted by the same symbol; its explicit form, on using eq. \eqref{cyclopvarphi}, is obtained to be
\begin{equation}\label{KMQsignrulelambda}
\lambda\phi(a_0,a_1,\ldots,a_p)=(-1)^{p
+|a_p|\sum\limits_{i=0}^{p-1}|a_i|}\phi(a_p,a_0,a_1,\ldots,a_{p-1}).
\end{equation}
Analogously, for the boundary operator we set
$b\,\phi(a_0da_1\cdots da_p)=\epsilon_p\,b\,\varphi(a_0da_1\cdots da_p)$; by means of eq. 
\eqref{boundopvarphi}, the computation yields
\begin{equation}\label{KMQsignruleb}
b\,\phi(a_0,a_1,\ldots,a_{p+1})=\sum_{i=0}^p(-1)^i\phi(a_0,\ldots,a_ia_{i+1},\ldots,a_{p+1})
+(-1)^{p+1+|a_{p+1}|\sum\limits_{j=0}^{p}|a_j|}\phi(a_{p+1}a_0,a_1,\ldots,a_{p}).
\end{equation}
For example, the latter form of the boundary operator on cochains is rather similar in appearance to that in the trivially graded case since it only differs by an additional contribution to the sign factor in the crossover  term; but the disadvantage in using the $\phi$s lies in the crucial fact that the simplicity in the definition $b\,\varphi=\varphi\circ b$ of the boundary operator for the $\varphi$s 
(see \eqref{defboundopgng}) then gets lost. 

\subsection{The graded commutative case}\label{Thegradedcommutativecase}

We shall have need for the graded commutative case; then a further reduction of $\Omega^1A$ is required 
(cf. also \cite{Kast88,Loda92}). This comes about from the additional relations
\begin{equation}\label{abeliancaserelonegng} 
db\,a=(-1)^{|a|(1+|b|)}a\,db
\end{equation}
that arise as follows. Let $A$ be a $\mathds{Z}/2$-graded algebra and $M$ a $\mathds{Z}/2$-graded left 
$A$-module with compatible gradings; if $A$ is graded commutative, we define a compatible right action by $v\,a=(-1)^{|a|\,|v|}a\,v$, which explains the above relation. On passing to $\Omega_{}^2A$, and also higher values of $p$, the computation
$$d^{\,2}(a\,b)=d(da\,b+(-1)^{|a|}a\,db)=d((-1)^{(1+|a|)|b|}b\,da+(-1)^{|a|}a\,db)=(-1)^{(1+|a|)|b|}db\,da+(-1)^{|a|}da\,db=0$$
shows that the further relations 
\begin{equation}\label{abeliancasereltwogng}
db\,da=(-1)^{(1+|a|)(1+|b|)}da\,db
\end{equation}
must be guaranteed; they demonstrate that in the trivially graded case the differentials anticommute, whereas in the nontrivially graded case for, e.g., $|a|=1=|b|$ they commute. The relations 
\eqref{abeliancaserelonegng} and \eqref{abeliancasereltwogng} must be respected on defining the universal 
(bi)graded differential algebra.

For this purpose, let us first consider the general non commutative $\mathds{Z}/2$-graded case; the elements $da$ of $\Omega^1A$ can be defined by
\begin{equation}\label{defofdgradedcasegng}
da=e\otimes a-(-1)^{|a|}a\otimes e
\end{equation}
on the tensor product $A\otimes A$, which is understood to be graded with the product $a\otimes b\cdot c\otimes d=(-1)^{|b|\,|c|}ac\otimes bd$; with this definition, the graded 
Leibniz rule is easily verified. Then $\Omega^1A$ may be characterized as the kernel of the modified 
multiplication map $\tilde{\mu}=\mu\circ(\text{id}\otimes\varepsilon)$ where $\varepsilon$ denotes the grading operator, i.e.
\begin{equation}\label{defmutildegng}
\tilde{\mu}(a\otimes b)=(-1)^{|b|}a\,b.
\end{equation}
Furthermore, the $A$-bimodule of $p$-forms is $\Omega^pA=\overset{p}{\otimes}^{}_A\Omega^1A$, where again the tensor product, now over $A$, is understood to be graded.

Suppose then $A$ is graded commutative; as to property \eqref{abeliancaserelonegng}, we observe that the following identity holds
\begin{equation*}
a\,db-(-1)^{|a|(1+|b|)}db\,a=(-1)^{|a|}(e\otimes a-(-1)^{|a|}a\otimes e)(e\otimes b-(-1)^{|b|}b\otimes e)
\end{equation*}
for the verification of which the gradation of the tensor product and graded commutativity is used. Hence we have $[a,db]\in(\Omega^1A)^2$, where $(\Omega^1A)^2$ is a subbimodule of the graded tensor product $A\otimes A$, and this is also contained in $\Omega^1A$ by construction. The above identity instructs us to pass to the quotient
\begin{equation}
\Omega^1(A):=\Omega^1A/(\Omega^1A)^2
\end{equation}
the elements of which thus obey $[a,db]=0$; again, the space of $1$-forms $\Omega^1(A)$ is distinguished by the universality property, as is easily verified.

Turning to higher values of  $p$ in the trivially graded commutative case, one can simply set 
\begin{equation}
\Omega^p(A)=\overset{p}{\wedge}^{}_A\Omega^1(A)
\end{equation}
and $\Omega^{\displaystyle\cdot}(A)=\underset{p\geq 0}{\oplus}\Omega^1(A)$; the skewsymmetric tensor product is as usual given by
\begin{equation}
a^{}_0da_1\wedge\cdots\wedge da_p=\frac{1}{\sqrt{p\,!}}\sum_{\sigma\in S_p}\varepsilon(\sigma)
a^{}_0da_{\sigma^{-1}(1)}\otimes\cdots\otimes da_{\sigma^{-1}(p)}
\end{equation}
where $\varepsilon(\sigma)$ denotes the sign of the permutation $\sigma$; with this definition, the relations \eqref{abeliancasereltwogng} are guaranteed. 

In the nontrivially graded commutative situation, however, the standard symmetrization does not work because the relations $db\,da=(-1)^{|da|\,|db|}da\,db$ depend on the parities. Hence, since  $\Omega^1(A)$ is  
$\mathds{Z}/2$-graded with $\varepsilon(a\,db)=(-1)^{|a|+|db|}a\,db$, a new concept is needed.

To begin with, consider a nontrivially graded vector space $V$ of dimension $n$ and its graded tensor product $V\otimes V$; from this we can construct the symmetrized graded symmetric tensor product $V\vee V$ and the graded antisymmetric tensor product $V\wedge V$ defined by
\begin{equation}
v\vee w=v\otimes w+(-1)^{|v|\,|w|}w\otimes v\qquad\qquad\qquad\qquad:\,v,w\in V
\end{equation}
and
\begin{equation}
v\wedge w=v\otimes w-(-1)^{|v|\,|w|}w\otimes v\qquad\qquad\qquad\qquad:\,v,w\in V
\end{equation}
on simple tensors, and extended to $V\otimes V$ through linearity. These products obey $w\vee v=(-1)^{|v|\,|w|}v\vee w$ and $w\wedge v=-(-1)^{|v|\,|w|}v\wedge w$, and so they are graded  commutative and graded anticommutative, respectively; in the trivially graded case, one regains the standard versions. For higher powers, the general construction is obvious now; the prefactor of a particular summand is the sign $(\pm1)^{|\sigma|}$ of the permutation $\sigma$, 
times the corresponding Koszul-Milnor-Quillen sign.

Applied to the case at hand, let $M$ be a $\mathds{Z}/2$-graded $A$-bimodule and $M\otimes^{}_AM$ the graded tensor product over $A$; we then have the product $M\vee^{}_AM$ available. The $A$-bimodule $\Omega^p(A)$ of $p$-forms
for a nontrivially graded commutative algebra $A$ can thus be defined as
\begin{equation}
\Omega^p(A)=\overset{p}{\vee}^{}_A\,\Omega^1(A).
\end{equation}
Its decomposable elements read as
\begin{equation}
a^{}_0\,da^{}_1\vee\cdots\vee da_p\qquad\qquad\qquad:\,a^{}_0,a^{}_1,\cdots,a_p\in A
\end{equation}
where again the tensor product over $A$ is understood.

\section{Berezin Integration and Graded Cyclic Cohomology}

In the present section we apply the general machinery, having been developed for the differential calculus and the integral calculi on a $\mathds{Z}/2$-graded algebra, to one particular example; this is the Gra{\ss}mann algebra, which figures in 
quantum field theories with fermions. What we aim at is to investigate whether Connes' characters over 
a Gra{\ss}mannn algebra have something to tell us about the Berezin integral calculus over anticommuting variables. If the question can be answered in the affirmative, then we expect that the general theory of Connes' characters should support (at least some of) the Berezin rules so that they are put on a firmer ground. If not, then this no go result would say, noncommutative geometry is not capable to include supersymmetric ideas; a disappointing result. As we shall demonstrate, however, Connes' approach is indeed capable to give us valuable insights into the origin of the Berezin integration rules; similar ideas were enounced by Kastler, but his introductory remark in
\cite{Kast88} does not go beyond the one-dimensional case, which is almost trivial.

\subsection{Differential calculus on the Gra{\ss}mann algebra}

Let us recall that the Gra{\ss}mann algebra, denoted $G^n$, is the associative unital algebra over the reals with $n$ generators $\xi^i$ and relations 
\begin{equation}\label{bcch1}
\xi^i\xi^j+\xi^j\xi^i=0\qquad\qquad\qquad\qquad:\, i,j\in\{1,\ldots,n\}.
\end{equation}
It is isomorphic to the exterior algebra over an $n$-dimensional real vector space, but this isomorphism is not natural; at any rate, for our purposes the definition through generators and relations is the one we need. As a $2^n$-dimensional basis of the underlying vector space one can choose the elements
\begin{equation}\label{bcch2}
\xi^{j_1}\cdots\xi^{j_q}\qquad\qquad\qquad\qquad:\,j_1< \cdots <j_q
\end{equation}
with $0\leq q\leq n$; the case $q=0$ is meant to signify the unit element, denoted by $1$ in the present case. The algebra $G^n$  is obviously $\mathds{Z}/2$-graded commutative, and so we must specialize the general theory of the last section to the graded commutative situation. Accordingly, the generators $\xi^i$ all have parity $|\xi^i|=1$, and the general element $f\in G^n$ can be written as
\begin{equation}\label{bcch3}
f(\xi)=\sum\limits_{q=0}^n \frac{1}{q\,!}\,f_{j_1\cdots j_q}\xi^{j_1}\cdots\xi^{j_q}
\end{equation}
with the coefficients $f_{j_1\cdots j_q}\in\mathds{R}$ being totally skew in their indices. 
We also endow the Gra{\ss}mann algebra with a $\ast$-structure, though $G^n$ is considered as an algebra over $\mathds{R}$; it is a linear involutive automorphism defined by $(\xi^i)^{\ast}=\xi^i$ and
\begin{equation}\label{bcch3a}
(\xi^{\,j_1}\cdots\xi^{\,j_q})^{\ast}=\xi^{\,j_q}\cdots\xi^{\,j_1}
\end{equation}
so that its only effect is to reverse the order of factors. Note that most authors use a different convention; first of all, the algebra $A$ is taken over $\mathds{C}$ so that the involution is antilinear, and in the $\mathds{Z}/2$-graded case one sets $(ab)^{\ast}=(-1)^{|a|\,|b|}b^{\ast}a^{\ast}$ for all $a,b\in A$. If this definition is extended to the real 
case, the $\ast$-involution acts as the identity for the case at hand. Our version, which is also used by DeWitt 
\cite{DeWi84}, will help to simplify many of the calculations to follow.

For Taylor expansion, one needs the finite difference $\delta f(\xi)=f(\xi+\delta\xi)-f(\xi)$, where $\xi+\delta\xi$ with
$\delta\xi^i=\epsilon\xi^i$ is an `infinitesimal' basis transformation, i.e. $\epsilon\ll 1$; whence, the $\delta\xi$ are odd quantities. For the computation it is convenient to introduce the formal partial derivatives 
\begin{equation}\label{bcch4}
\frac{\partial}{\partial\xi^j}(\xi^{j_1}\cdots\xi^{j_q})=
\sum_{r=1}^q(-1)^{r-1}\xi^{j_1}\cdots\delta^{j_r}{}_{j}\cdots\xi^{j_q}.
\end{equation}
which are elements of $\text{Der}(G^n,G^n)$ of  parity $|\partial_i|=1$, and one obtains $\delta=\delta\xi^{j}\,\partial/\partial\xi^j$, which is an even operator.

Let us turn to general elements $X$ of $\text{Der}(G^n,G^n)$; they are all outer derivations since $G^n$ 
is graded commutative. For a proper treatment one needs translations, and for these one must introduce a 
second Gra{\ss}mann algebra $G^n$ with the same number of generators, denoted by $\eta^i$; on introducing 
the graded tensor product $G^n\otimes G^n$, we can identify the $\eta$s and $\xi$s with the elements 
$\eta^i\otimes 1$ and $1\otimes\xi^i$, respectively, and the graded tensor product guarantees that they anticommute.
In this way it makes sense to consider the translated generators $\xi^{\,\prime i}=\xi^i+s^i\eta^i$ with $s^i\in\mathds{R}$, and also  
the translated element $f(\xi^{\,\prime})\in G^n\otimes G^n$, which is manipulated as follows
\begin{align*}
f(\xi^{\,\prime})&=f(\xi)+\int\limits_0^1dt\,\frac{\partial}{\partial t}\,f(\xi+t(\xi-\xi^{\,\prime}))\\
&=f(\xi)+(\xi-\xi^{\,\prime})^i\int\limits_0^1dt\,(\partial_if)(\xi+t(\xi-\xi^{\,\prime}))\\
&=f(\xi)+(\xi-\xi^{\,\prime})^i\,g^{}_i(\xi^{\,\prime}).
\end{align*}
We thus obtain
$$(Xf)(\xi^{\,\prime})=X\xi^{\,\prime i}\,g^{}_i\,(\xi^{\,\prime})+(-1)^{|X|}
(\xi-\xi^{\,\prime})^iXg^{}_i(\xi^{\,\prime})$$
and letting $\xi^{\,\prime}\to\xi$, we arrive at
\begin{equation}\label{bcch5}
(Xf)(\xi)=X^i(\xi)\partial_if(\xi)\qquad\qquad\qquad\qquad:\,X^i(\xi)=X(\xi^i).
\end{equation}
We thus have proven that the derivations $X\in\text{Der}(G^n,G^n)$, called Gra{\ss}mann vector fields, form a free $\mathds{R}$-module of rank $n$, with the partial derivatives $\partial_i$ as basis. Since the Gra{\ss}mann algebra is graded commutative, the derivations $X$ form a left $G^n$-module, denoted by $V(G^n)$, which can be given the structure of a $\mathds{Z}/2$-graded Lie algebra.

Finite transformations of Gra{\ss}mann variables are obtained on exponentiating vector fields $X$ of parity $|X|=0$, i.e. all the components $X^i$ must be odd. Since the ordinary (ungraded) commutator obeys $[X,X]=0$, the standard property $e^{X}e^{-X}=e^{-X}e^{X}=1$ of the exponential remains intact, and so the transformation
$$\xi^{\prime i}=e^{X(\xi)}\xi^{i}=\xi^{i}+X^i(\xi)+\frac{1}{2}\,X^j(\xi)\partial_jX^{i}(\xi)+\cdots$$
is invertible. Special cases are translations with the vector field
\begin{equation}\label{bcch7}
X_{\eta}(\xi)=\eta^i\frac{\partial}{\partial\xi^i}\qquad\qquad\qquad:\,e^{X_{\eta}(\xi)}\xi=\xi+\eta
\end{equation}
where the $\eta$s are the generators of a second $G^n$ as discussed above, and for a homogeneous transformation the generating vector field is 
\begin{equation}\label{bcch8}
X_{\alpha}(\xi)=\alpha^{i}{}_j\xi^j\frac{\partial}{\partial\xi^i}\qquad\qquad\qquad:\,e^{X_{\alpha}(\xi)}\xi^i=A^i{}_j\xi^j
\end{equation}
with $\alpha^i{}_j\in\mathds{R}$ and $A(\alpha)=\exp\alpha$ an element of the general linear group in $n$ dimensions. In particular, for $\alpha^{i}{}_j=\lambda\,\delta^{i}{}_j$ one has 
\begin{equation}\label{bcch9}
X_{\lambda}(\xi)=\lambda\,\xi^i\frac{\partial}{\partial\xi^i}\qquad\qquad\qquad:
\,e^{X_{\lambda}(\xi)}\xi^{i}=e^{\lambda}\xi^{i}
\end{equation}
so that one obtains the Gra{\ss}mann analogue of a dilatation; this was used above for the Taylor expansion.

The importance of the affine transformations $\xi\mapsto A\xi+\eta$ resides in the fact that they constitute the automorphism group of the defining relations, i.e. the property \eqref{bcch1} remains intact.

\subsection{Berezin integration}

The discussions of the Berezin integration rules on a Gra{\ss}mann algebra as given in the physical literature are mostly heuristic; if not, they are more or less oriented towards the supersymmetric situation (see \cite{Bere87} and \cite{Leit80,DeWi84,Vlad84,Cons94,Deli99}). In the mathematical literature, 
treatments of this theme are rare; we are aware of \cite{Guil99}. The approach given below is related, but different from the one described in  
\cite{Vlad84}.

A priori, a Berezin `integral' is an $\mathds{R}$-linear map $J:G^n\to\mathds{R}$ only; as such, it is an element of the algebraic dual $(G^n)^{\prime}$, and so the latter linear space has the same dimension as the original $G^n$ itself. The multitude of possible maps $J$ is further narrowed down by the request for translational invariance. So let $s\eta^i$ denote a translation of the generators $\xi^i$ of $G^n$, as explained in the previous paragraph; for notational simplicity, the additional dependence on the real parameter $s$ is suppressed. For $f\in G^n$, the translated algebra element is then defined by
$$T(\eta)f(\xi)=f(\xi-\eta).$$
A Berezin `integral' $J$ is said to be translational invariant if the condition $J(T(\eta)f)=J(f)$ is fulfilled, where both $f$ and the translation are arbitrary. Actually, such a condition makes not much  sense since the left hand side is $\eta$-valued, and we have not yet defined what this means; the 
correct version will be given below.

For this purpose, let $A$ and $B$ be two $\mathds{Z}/2$-graded algebras, and $A\otimes B$ their graded tensor product; the latter may be viewed as a left $A$-module under the action $a^{\prime}(a\otimes b)=a^{\prime}a\otimes b$, and also $A$ may trivially be regarded as a left $A$-module. For the case at hand, $A=G^n$ is generated by the $\eta$s and $B=G^n$ by the $\xi$s, the original generators.
Consider then an $A$-linear map $J^{\,\prime}:A\otimes B\to A$; this means that $J^{\,\prime}(a\otimes b)=a\,J^{\,\prime}(e^{}_A\otimes b)$ and so $J^{\,\prime}$ is uniquely determined by the values it takes on $B$. Also observe that $J^{\,\prime}|_{B}:B\to A$ is an $\mathds{R}$-linear map with values in $A$. We now impose the further condition 
$$J^{\,\prime}(e^{}_A\otimes b)=e^{}_A\,J(b)$$
where $J:A\to\mathds{R}$ is a given $\mathds{R}$-linear map, which thus determines $J^{\,\prime}$ uniquely. A proper definition of the invariance condition then reads
\begin{equation}\label{bcch9a}
J^{\,\prime}(T(\eta)f)=J(f)
\end{equation}
which is a further condition on the admissible maps $J^{\,\prime}$, saying that $J^{\,\prime}$ is independent of translations.

We now claim, the above invariance requirement \eqref{bcch9a} determines the Berezin `integral' $J$ 
uniquely; the proof is straightforward. By means of the expansion 
$$f(\xi)=\sum_If^{}_I\xi_{}^I$$
where the $\xi_{}^I$ denote the basis with $I$ a (lexicographically ordered) multi-index, we have
$$J^{\,\prime}(T(\eta)f)=\sum_{I,K\atop |I|+|K|\leq n}\eta_{}^Kf^{}_{K,I}J(\xi_{}^I)$$
on using $G^n$-linearity with respect to the first factor. Consequently, all terms with $|I|+|K|< n$ must vanish, and for $|I|+|K|=n$ only the term with $K=0$ can survive. In total, translational invariance enforces $J(\xi_{}^I)=0$ for $I<n$, or   
\begin{equation}\label{bcch9b}
J(\xi^{i_1}\cdots\xi^{i_p})\,=\,\begin{cases}0&:\,p<n\\\varepsilon^{i_1\cdots i_n}&:\,p=n\end{cases}
\end{equation}
with a suitable normalization; an equivalent form is
\begin{equation}\label{bcch9b'}
J(f)\,=\,\frac{\partial^n}{\partial\xi^n\cdots\partial\xi^1}f(\xi)=f_{1\cdots n}.
\end{equation}
For a homogeneous transformation $T(A)f(\xi)=f(A^{-1}\xi)$ with $A$ a real $n\times n$-matrix, this entails the transformation law
\begin{equation}\label{bcch9b''}
J(T(A)f)\,=\,\text{det}(A)^{-1}J(f).
\end{equation}
The defining property \eqref{bcch9b'} and its consequence \eqref{bcch9b''} are the only Berezin integration rules that we accept as valid.

Recall, the map $J$ is simply a special element of the algebraic dual of $G^n$; but it has become customary, and this is the weak point, to rewrite it as a kind of integral:
\begin{equation}\label{bcch9c}
J(f)=\int d^{\,n}\xi\,f(\xi).
\end{equation}
This rewriting, however, is void of any definite meaning since a precise definition of the `volume element' $d^{\,n}\xi$ is missing. It is the ultimate purpose of the present paper to provide for a rigorous justification of this formal notation.

\subsection{Fourier transformation and inner products on the Gra{\ss}mann algebra}

Before turning to the problem just raised, some more technique is needed. In what follows we shall make use of the notational convention \eqref{bcch9c}, which simply amounts to the instruction to sort out the coefficient $f_{1\cdots n}$ of the term of highest degree, and no more. As an immediate consequence of the rules \eqref{bcch9b} one obtains the further property
\begin{equation}\label{bcch9d}
\int d^{\,n}\xi\,\partial_if(\xi)=0
\end{equation}
saying that a `boundary integral' vanishes. We shall also have need for Fourier transformation, defined by
\begin{equation}\label{bcch9e}
Ff(\xi)=\int d^{\,n}\xi^{\,\prime}\,e^{\xi\cdot\xi^{\,\prime}}f(\xi^{\,\prime})
\end{equation}
where $\xi\cdot\xi^{\,\prime}=\xi^i\delta_{ij}\xi^{\,\prime j}$ with the $\xi^{\,\prime}$s the generators of a further $G^n$; hence, the underlying structure is again the graded tensor product $G^n\otimes G^n$ so that the two sets of generators anticommute. Fourier transformation obeys $F^2=(-1)^{{n+1\choose 2}}\,1$; furthermore, it is related to the conventional Hodge dual, as follows from the explicit form
\begin{equation}\label{bcch9f}
Ff(\xi)=\sum_q\frac{1}{\bar{q}\,!}\,(-1)^{\bar{q}+{\bar{q}\choose 2}}\frac{1}{q\,!}f_{j_1\cdots j_q}\varepsilon^{j_1\cdots j_qj_{q+1}\cdots j_n}\xi_{j_{q+1}}\cdots\xi_{j_n}
\end{equation}
where $\bar{q}=n-q$ for short, and $\varepsilon^{j_1\cdots j_n}$ the Levi-Civita symbol. We are now ready to introduce an inner product on $G^n$, viewed as a linear space, defined by
\begin{equation*}
(f|g)=(-1)^{{n\choose 2}}\int d^{\,n}\xi^{\,\prime}(Ff^{\ast})(-\xi^{\,\prime})\,g(\xi^{\,\prime})
\end{equation*}
in which also the algebra structure enters. In the form of a twofold integral it reads 
\begin{equation}\label{bcch9g}
(f|g)=(-1)^{{n\choose 2}}\int d^{\,n}\xi^{\,\prime}\int d^{\,n}\xi\,e^{\,\xi\cdot\xi^{\,\prime}}
\,f(\xi)^{\ast}\,g(\xi^{\,\prime})
\end{equation}
and its explicit version is obtained to be
\begin{equation}\label{bcch9h}
(f|g)=\sum_{q=0}^n\frac{1}{q\,!}\,\bar{f}_{j_1\cdots j_q}\,g^{j_1\cdots j_q}
\end{equation}
where the raising and lowering of indices is performed with the Kronecker metric. Since in the present context the complexification of $G^n$ is considered, complex conjugation denoted by an overbar gets involved. As is obvious now, the inner product \eqref{bcch9g} is both 
hermitean and positive definite. 

But it is not the only inner product we shall have need for; another even more natural one is obtained on simply defining
\begin{equation}\label{bcch9i}
\langle f|g\rangle=i^{{n\choose 2}}\int d^{\,n}\xi\,f(\xi)^{\ast}\,g(\xi).
\end{equation}
It is nondegenerate, and the prefactor is designed so as to guarantee hermiticity, as follows from the explicit form 
\begin{equation}\label{bcch9j}
\langle f|g\rangle=i^{{n\choose 2}}\sum_{q=0}^n\frac{(-1)^{{q\choose 2}}}{q\,!\,\bar{q}\,!}\,
\bar{f}_{j_1\cdots j_q}\,\varepsilon^{j_1\cdots j_qj_{q+1}\cdots j_n}g_{j_{q+1}\cdots j_n}.
\end{equation}
But as opposed to the former, the present inner product, though being nondegenerate, is indefinite.  

Both these structures can be subsumed as follows under a common heading. Let us define a grading operator, which should not be confused with the $\mathds{Z}/2$-grading induced by the $\mathds{N}$-grading, through 
\begin{equation}\label{bcch9k}
(\gamma f_q)^{j_{q+1}\cdots j_n}=(-i)^{{n\choose 2}}\,\frac{(-1)^{\bar{q}+{\bar{q}\choose 2}}}{q\,!}\, 
\varepsilon^{j_{q+1}\cdots j_nj_1\cdots j_q}f_{j_1\cdots j_q}
\end{equation}
or, using the Hodge dual, by
\begin{equation}\label{bcch9l}
(\gamma f_q)(\xi)=(-i)^{{n\choose 2}}\,(-1)^{{q\choose 2}}(\ast f_q)(\xi).
\end{equation}
As is straightforward to verify, it indeed obeys $\gamma_{}^2=1$; beyond this, it is unitary with respect to the positive definite inner product \eqref{bcch9g}, i.e.
\begin{equation}\label{bcch9m}
(\gamma f|\gamma g)=(f|g).
\end{equation}
The same property is valid for the indefinite inner product \eqref{bcch9i}, which may now be written by means of the grading operator in terms of the positive definite inner product as
\begin{equation}\label{bcch9n}
\langle f|g\rangle=(f|\gamma g).
\end{equation}
Hence, through the inner product $(\,\cdot\,|\,\cdot\,)$ the Gra{\ss}mann algebra $G^n$ can be equipped with a Hilbert space structure; on the other hand, the inner product $\langle\,\cdot\,|\,\cdot\,\rangle$ is indefinite, which becomes manifest in the orthogonal decomposition $G^n=(G^n)_{+}\oplus(G^n)_{-}$ with $\gamma f_{\pm}=\pm f_{\pm}$ since
\begin{equation*}
\langle f|g\rangle=(f_{+}| g_{+})-(f_{-}| g_{-}).
\end{equation*}
The above structure, namely, a separable Hilbert space on which a unitary grading operator is defined that gives rise for an indefinite inner product according to eq. \eqref{bcch9n}, is sometimes called a 
Krein space (see \cite{Bogn74}). It underlies the construction of an even Fredholm module \cite{Conn94}, and also makes its appearance in the context of ghost fermions \cite{Gren02} which get involved in the Batalin-Vilkovisky approach to the Becchi-Rouet-Stora-Tyutin quantization of gauge theories (see, e.g., \cite{Henn92}). 

We conclude with the following observation. Let us introduce the operators $\eta^i=\xi^i$ and $\zeta_i=\partial_i$, which both act from the left. They obey the anticommutation relations
\begin{equation}\label{bcch9o}
[\zeta_i,\eta^j]_{+}=\delta_{i}{}^j
\end{equation}
and may also be viewed as the operators of external and internal multiplication. With respect to the inner product $\langle\,\cdot\,|\,\cdot\,\rangle$, these operators are selfadjoint, i.e.
\begin{equation}\label{bcch9p}
\langle f|\zeta_ig\rangle=\langle\zeta_i f|g\rangle\qquad\qquad\qquad\qquad
\langle f|\eta^ig\rangle=\langle\eta^i f|g\rangle.
\end{equation}
This property relies on the identity 
\begin{equation}\label{bcch9q}
\langle f|\partial_ig\rangle=\langle f|\partial_ig\rangle
\end{equation}
being valid for any elements $f,g\in G^n$, i. e. without any restriction to homogeneous elements.
By contrast, with respect to the inner product $(\,\cdot\,|\,\cdot\,)$, the operators $\eta^i=\xi^i$ and $\zeta_i=\partial_i$ are adjoint to one another, i.e.
\begin{equation}\label{bcch9r}
(f|\zeta^ig)=(\eta^if|g)\qquad\qquad\qquad\qquad
\langle f|\eta^ig\rangle=\langle\zeta^i f|g\rangle
\end{equation}
where $\zeta^i=\delta^{ij}\zeta_j$. In the latter case, we introduce the further operators
\begin{equation}\label{bcch9s}
\gamma^i=\eta^i+\delta^{ij}\zeta_j
\end{equation}
which form a Clifford algebra, i.e. $\gamma^i\,\gamma^j+\gamma^j\,\gamma^i=2\delta^{ij}$; also, the $\gamma$s are selfadjoint operators. Restricting ourselves to the case of an even $n$, the  representation of the $\gamma$s on the (complexified) Gra{\ss}mann algebra is not irreducible; it is equivalent to the Dirac representation. But it acts irreducibly on the subspace of even elements. For the indefinite inner product $\langle\,\cdot\,|\,\cdot\,\rangle$, we choose 
\begin{equation}\label{bcch9t}
\gamma^i=\zeta^i\qquad\qquad\qquad\gamma^{n+i}=\eta^i
\end{equation}
with
\begin{equation}\label{bcch9u}
([\gamma^a,\gamma^b]_{+})_{a,b=1,\ldots,2n}=\begin{pmatrix} & 1_n\\1_n& \end{pmatrix}.
\end{equation}
They obey $\langle f|\gamma^ag\rangle=\langle f|\gamma^ag\rangle$ and form a Clifford algebra with $2n$ generators, with an indefinite metric of signature zero; in this latter case, the representation is irreducible.

Finally, returning to the Clifford algebra with generators \eqref{bcch9s}, we can devise a Dirac type operator
\begin{equation}\label{bcch9x}
D=\gamma^i\frac{\partial}{\partial\xi^i}
\end{equation}
which reduces to $D=(\xi^i+\delta^{ij}\partial_j)\partial^{}_i=\xi^i\partial^{}_i$  and thus coincides 
(cf. \eqref{bcch9}) with the Gra{\ss}mannian version of Euler's dilatation operator.

\subsection{The universal differential bigraded Gra{\ss}mann algebra}

Let us return to the problem raised in the last but one subsection, namely, whether the `volume element' 
$d^{\,n}\xi$ that enters the formal Berezin integral, can be given a precise meaning. We approach this 
problem on taking recourse to the general construction of the universal differential graded algebra of a 
general noncommutative algebra; this will supply for the necessary abstract tools to investigate these questions.  

We have discussed the universal differential calculus for a graded commutative algebra $A$ in Subsec. 
\ref{Thegradedcommutativecase}; 
the results obtained there are now applied to the situation where $A$ is defined through generators and relations, as it is the case for the Gra{\ss}mann algebra. 

As we know, in the graded commutative case one must pass to the $G^n$-bimodule $\Omega_{}^1(G^n)$ of 
$1$-forms. This is a free module of rank $n$, for which the differentials $d\xi^i$ form basis; their parity is zero. As should be noted, there is no conflict with the fact that finite differences have parity one; the  $\delta\xi^i$ and the differentials $d\xi^i$ are essentially different constructs. Differential forms of order $p$ are obtained on forming the graded symmetrized product $\Omega_{}^p(G^n)=\overset{p}{\vee}_{G^n}\Omega_{}^1(G^n)$ so that the $d\xi^i$ commute:
\begin{equation}\label{bcch10}
d\xi^i\,d\xi^j=d\xi^j\,d\xi^i.
\end{equation}
This assignment guarantees the operator of exterior differentiation to have parity $|d|=1$;
for $f\in G^n$ we then obtain
$$d^{\,2}f(\xi)=d(\partial_jf\,d\xi^j)=\partial_i\partial_jf\,d\xi^id\xi^j=0$$
since the partial differentiations anticommute and the differentials commute, which is the result one wants. Hence, for the construction of the basis for the module of $p$-forms we must select the graded 
symmetrized tensor product
\begin{equation}\label{bcch11}
d\xi^{i_1}\vee\cdots\vee d\xi^{i_p}=\frac{1}{\sqrt{p\,!}}\,\sum_{\sigma\in S_n}
d\xi^{i_{\sigma(1)}}\otimes\cdots\otimes d\xi^{i_{\sigma(p)}}
\end{equation}
that here coincides with the standard symmetrization. But the graded version is needed for a general 
$p$-form, which is a finite sum
\begin{equation}\label{bcch12}
\omega_p=\sum f^{}_0\,df^{}_1\vee\cdots\vee df_p
\end{equation}
where $f_i\in G^n$; the symbol $\vee$ signifying the symmetrization will often be suppressed. We thus can write down the general $p$-form
\begin{equation}\label{bcch13}
\omega_p(\xi)=\frac{1}{p\,!}\,\omega_{i_1\cdots i_p}(\xi)\,d\xi^{i_1}\vee\cdots\vee d\xi^{i_p}
\end{equation}
with
\begin{equation}\label{bcch14}
\omega^{}_{i_1\cdots i_p}(\xi)=\frac{1}{q\,!}\,\omega_{i^{}_1\cdots i_p,j^{}_1\cdots j_q}\,\xi^{j^{}_1}\cdots 
\xi^{j_q}
\end{equation}
where $\omega_{i^{}_1\cdots i_p,j^{}_1\cdots j_q}$ is completely symmetric in the indices 
$i^{}_1,\ldots,i_p$, and completely antisymmetric in the indices $j^{}_1\ldots,j_q$; furthermore, the operator of exterior differentiation $d$ acts on $\omega_p$ via
\begin{equation}\label{bcch15}
d\omega_p(\xi)=\frac{1}{p\,!}\partial_{i_0}\omega_{i_1\cdots i_p}(\xi)\,d\xi^{i_0}\vee d\xi^{i_1}\vee\cdots\vee d\xi^{i_p}
\end{equation}
and obeys $d\circ d=0$ by construction. To resume, we now have available the universal differential algebra $\Omega^{\displaystyle\cdot}(G^n)$; the general element of $\Omega^p(G^n)$ may be written as a finite sum of terms $f_0\,df_1\vee\cdots\vee df_p$ with parity 
\begin{equation}\label{bcch16}
|f_0\,df_1\vee\cdots\vee df_p|=p+\sum_{i=0}^p|f_p|.
\end{equation}
Furthermore, $\Omega^{\displaystyle\cdot}(G^n)$ is graded commutative, i.e. $[\omega^{}_{p},\omega^{\,\prime}_{p^{\,\prime}}]=0$ for all $\omega^{}_{p}\in 
\Omega^p(G^n)$ and $\omega^{\,\prime}_{p^{\,\prime}}\in 
\Omega^{p^{\,\prime}}(G^{n})$, since $G^n$ is. 

Let us compare the present situation with that of conventional differential forms over a smooth manifold $M$; in this case, where
$$\omega_p(x)=\frac{1}{p\,!}\omega_{i_1\cdots i_p}(x)\,dx^{i_1}\wedge\cdots\wedge dx^{i_p}$$
with $x\in M$, the coefficient functions $\omega_{i_1\cdots i_p}(x)$ contain arbitrary powers of the $x^i$, whereas the  $\omega_p$ vanish for all $p>n$ because the differentials anticommute. By contrast, in the 
Gra{\ss}mann case 
the expansion of the coefficient functions terminate for powers higher than $n$, but there is no restriction on the degree $p$ of the differential forms since the differentials $d\xi^i$ commute. 

We have already discussed the left $G^n$-module $V(G^n)$ of vector fields, which is free of rank $n$, with the partial derivatives $\partial_i$ forming a basis. Also in this case we can form tensor products, which we denote by $\Omega_p(G^n)$, with $\Omega^{}_0(G^n)=G^n$ and $\Omega^{}_1(G^n)=V(G^n)$; but before doing so we must specify the parity of the partial derivatives. We expect the relevant construction to be the graded symmetrization $\Omega_p(G^n)=\overset{p}{\vee}_{G^n}\Omega_1(G^n)$; in order to have $\partial_i\vee\partial_j=\partial_j\vee\partial_i$, this requirement enforces the assignment $|\partial_i|=0$, contrary to naive expectation, so that 
\begin{equation}\label{bcch17}
\frac{\partial}{\partial\xi^{i_1}}\vee\cdots\vee\frac{\partial}{\partial\xi^{i_p}}=\frac{1}{\sqrt{p\,!}}\,\sum_{\sigma\in S_n}
\frac{\partial}{\partial\xi^{i_{\sigma(1)}}}\otimes\cdots\otimes
\frac{\partial}{\partial\xi^{i_{\sigma(p)}}}.
\end{equation}
For the general element $\chi^p\in\Omega_p{G^n}$, called a Gra{\ss}mann \emph{current}, we are then guaranteed that the $G^n$-valued tensor components $\chi^{i^{}_1\cdots i_p}$ in the expansion
\begin{equation}\label{bcch18}
\chi^p=\frac{1}{p\,!}\,\chi^{i^{}_1\cdots i_p}\partial_{i^{}_1}\vee\cdots\vee\partial_{i_p}
\end{equation}
are symmetric in their indices.

The construction of the module of $p$-forms $\Omega^p(G^n)$ and the space of $p$-currents $\Omega_p(G^n)$ is quite different; it is for that reason why we have not defined the differentials $d\xi^i$ as the dual basis of the $\partial_i$, the partial derivatives. After the fact, however, we can define a pairing between these spaces. For this, let $A$ be an involutive $\mathds{Z}/2$-graded algebra; given  two left $A$-modules $M$ and $N$, both being equipped with compatible $\mathds{Z}/2$-gradings, we define an 
A-valued pairing of $M$ and $N$ as a sesquilinear map $\langle\,\cdot\,|\,\cdot\,\rangle:M\times N\to A$, subject to the condition
\begin{equation}\label{bcch19}
\langle av|bw\rangle=a^{\ast}\langle v|w\rangle b\qquad\qquad\qquad:\,a,b\in A;\,\,v\in M,\,\,w\in N.
\end{equation}
In case that $M=N$, one can then define a hermitean structure (see, e.g. \cite{Grac01}). 
For the construction of a $G^n$-valued pairing $\langle\,\cdot\,|\,\cdot\,\rangle:\Omega_p(G^n)\times\Omega^p(G^n)\to G^n$, we begin with $p=1$ and set 
$\langle \partial_i|d\xi^j\rangle=\delta_i{}^j$ by definition.
The extension to values of $p\geq 1$ is thus obtained to be 
\begin{equation}\label{bcch20}
\langle\frac{\partial}{\partial\xi^{i_1}}\vee\cdots\vee\frac{\partial}{\partial\xi^{i_p}}|
\,d\xi^{j_1}\vee\cdots\vee d\xi^{j_q}\rangle=\delta_p{}^q\sum_{\sigma}\delta_{i_{\sigma(1)}}{}^{j_1}\cdots
\delta_{i_{\sigma(p)}}{}^{j_p}
\end{equation}
where the normalization made in the eqs. \eqref{bcch11} and \eqref{bcch17} proves to be essential. In this way, the $G^n$-valued pairing is given by
\begin{equation}\label{bcch21}
\langle\chi^p(\xi)|\omega_p(\xi)\rangle=\frac{1}{p\,!}\chi^{i_1\cdots i_p}(\xi)^\ast\,
\omega_{i_1\cdots i_p}(\xi)
\end{equation}
which can be made a nondegenerate $\mathds{R}$-valued pairing $\langle\,\cdot\,|\,\cdot\,\rangle:\Omega_p(G^n)\times\Omega^p(G^n)\to \mathds{R}$ by means of the Berezin integral:
\begin{equation}\label{bcch22}
\langle\chi^p|\omega_p\rangle=\int d^{\,n}\xi\,\langle\chi^p(\xi)|\omega_p(\xi)\rangle.
\end{equation}
As opposed to the inner product \eqref{bcch9i}, we here omit the prefactor since from now on $G^n$ is considered as an algebra over the reals. The basic identity \eqref{bcch9q}, i.e. $\langle\chi^p|\partial_i\omega_p\rangle=\langle\partial_i\chi^p|\omega_p\rangle$ remains valid, by means of which many of the computations to follow are drastically simplified. Hence, the space of $p$-currents is the dual of the space of $p$-forms. 

In addition, the space of $p$-forms $\Omega^p(G^n)$ can be equipped with a positive definite inner product. For this, we introduce a metric on $\Omega^p(G_{}^1)$ by
\begin{equation}\label{bcch23}
\langle d\xi^i|d\xi^j\rangle=\delta^{ij}
\end{equation}
with the extension to the basis of $\Omega^p(G^n)$ being obtained by similar manipulations as above; in this way, we can freely raise and lower indices. We now define the scalar product by
\begin{equation}\label{bcch24}
(\omega_p|\lambda_p)=(-1)^{{n\choose 2}}\int d^{\,n}\xi^{\,\prime}\int d^{\,n}\xi\,
e^{\,\xi\cdot\xi^{\,\prime}}
\frac{1}{p\,!}\,\chi^{i_1\cdots i_p}(\xi)^\ast\,
\omega_{i_1\cdots i_p}(\xi^{\,\prime})
\end{equation}
for $\omega_p,\lambda_p\in\Omega^p(G^n)$, which is to be compared with eq. \eqref{bcch9g} above. Note that the decoration with additional indices does not alter the conclusions drawn there, but the relation \eqref{bcch9m} is no longer available.

The pairing and the inner product give rise to two operations, deriving from the operator \eqref{bcch15}  of exterior differentiation. Beginning with the pairing \eqref{bcch22}, we define the dual $\bar{d}$ of $d$ by
\begin{equation}\label{bcch25}
\langle\bar{d}\chi^p|\omega_p\rangle=\langle\chi^p|d\omega_p\rangle.
\end{equation}
which is calculated to be
\begin{equation}\label{bcch26}
\bar{d}\chi^p(\xi)=\frac{1}{(p-1)!}\,\partial_{i_1}\chi^{i^{}_1i^{}_2\dots i_p}(\xi)
\,\partial_{i^{}_2}\vee\cdots\vee \partial_{i_p}.
\end{equation}
This operator, a divergence which is also denoted below as $\partial=\bar{d}$, obeys $\bar{d}^{\,2}=0$ and is of degree minus one.
Turning to the inner product \eqref{bcch24}, the adjoint $d^{\ast}$ of $d$ is defined by
\begin{equation}\label{bcch28}
(\omega_{p}|d\lambda_{p-1})=(d^{\ast}\omega_{p}|\lambda_{p-1})
\end{equation}
and obeys $(d^{\ast})^2=0$; the calculation yields
\begin{equation}\label{bcch29}
d^{\ast}\omega_{p}(\xi)=\frac{1}{(p-1)!}\,\xi^{i^{}_1}
\omega_{i^{}_1i^{}_2\cdots i_p}(\xi)\,d\xi^{i^{}_2}\vee\cdots\vee d\xi^{i_p}.
\end{equation}
We write in the form of a contraction, i.e. $\iota^{}_D=d^{\ast}$, with $D=\xi^i\partial/\partial\xi^i$ the Euler vector field generating dilatations. For $X$ a general vector field of zero parity, let us introduce the contraction
\begin{equation}\label{bcch31}
\iota^{}_X\omega_{p+1}(\xi)=\frac{1}{p\,!}\,X^{i^{}_0}\omega_{i^{}_0i^{}_1\cdots i_p}(\xi)
\,d\xi^{i^{}_1}\vee\cdots\vee d\xi^{i_p}.
\end{equation}
We then define the Lie derivative $L^{}_X$ by the Cartan formula $L^{}_X=\iota^{}_Xd+d\iota^{}_X$, 
the explicit form of which is 
\begin{equation}\label{bcch32}
L^{}_X\omega_{p}=\left(\frac{1}{p\,!}\,X^{i}\partial_i\,\omega_{i^{}_1\cdots i_p}+
\frac{1}{(p-1)!}\,\partial_{i^{}_1}X^i\omega_{ii^{}_2\dots i_p}\right)
d\xi^{i^{}_1}\vee\cdots\vee d\xi^{i_p}.
\end{equation}
In particular, for the Euler vector field $D$ this gives
\begin{equation}\label{bcch33}
L^{}_D\omega_{p}=\left(D+p\right)\omega_{p}.
\end{equation}
In the conventional de Rham case, the Cartan formula is instrumental for a transparent proof of the Poincar\' e lemma (see \cite{Bott82}); this is as well the case in the present situation. The strategy of the proof consists in extracting a homotopy operator from the Cartan formula; the calculation is almost the same as in the classical situation and yields that an exact $p$-form $\omega_p$ can be written as $\omega_p=d\alpha_{p-1}$, with
\begin{equation}\label{bcch34}\alpha_{p-1}(\xi)=\int_0^1\,dt\frac{t^{p-1}}{(p-1)!}
\,\xi^{i^{}_1}\,\omega_{i^{}_1i^{}_2\cdots i_p}(t\xi)\,d\xi^{i^{}_2}\vee\cdots\vee d\xi^{i_p}.
\end{equation}
Whence, the complex $(\Omega^{\displaystyle\cdot}(G^n),d)$ is acyclic.

Observe that the classical case over the manifold $\mathds{R}^n$ and the Grassmann case have some similarities, but are also essentially different. This can be seen on having a closer look at the operator $\bar{d}$, defined in \eqref{bcch26}, which comes from the pairing \eqref{bcch22}; it can also be interpreted as acting on $\Omega^p(G^n)$ by means of the metric. Together with $d$ it gives rise to a rather unique candidate for a Laplacean type of operator, namely the square 
of the Dirac-Hodge-K\"ahler like operator $\not{\!d}=\bar{d}+d$; but  $\bar{d}\,d+d\,\bar{d}$ vanishes since it is equal to 
$g^{ij}\partial_i\partial_j$ so that in the 
Gra{\ss}mann case all $p$-forms are trivially harmonic. Turning to the positive definite inner product 
\eqref{bcch24} with the adjoint operator $d^{\ast}$, it permits to introduce the quadratic operator  $d^{\ast}d+d\,d^{\ast}$; as we have seen, however, this is the Lie derivative $L^{}_D$, which has not much in common with a Laplace operator. Nevertheless, we can prove the following analogue of the Hodge decomposition theorem
$$\Omega^{p}(G^n)=d\,\Omega^{p-1}(G^n)\oplus d^{\ast}\Omega^{p+1}(G^n)$$
where, similar to the classical situation over the flat manifold $\mathds{R}^n$, the contribution of the harmonic forms is absent. However, whereas the conventional proof requires elliptic operator theory 
\cite{Tayl96}, it is purely algebraic in the present case and follows standard arguments, which we omit.

In concluding this section, let us note that one can also define an external multiplication on 
$p$-forms by $\varepsilon(d\xi^i)=d\xi^i\vee$ which together with the internal multiplication $(\iota^{}_X=\iota(X))$ operator $\iota(\partial/\partial\xi^i)$ obey the commutation relations
$$[\iota\left(\frac{\partial}{\partial\xi^i}\right),\varepsilon(d\xi^j)]^{}_{-}=\delta_i{}^j.$$
Hence, $\iota(\partial/\partial\xi^i)$ and $\varepsilon(d\xi^i)$ may be viewed as bosonic 
annihilation and creation operators; this is just the opposite situation as compared to the de Rham case, where the analogues $\iota(\partial/\partial x^i)$ and $\varepsilon(dx^i)$ are fermionic operators, with 
$[\iota(\partial/\partial x^i),\varepsilon(dx^j)]^{}_{+}=\delta_i{}^j$ an anticommutator. 

\section{Connes' integral calculi on a Gra{\ss}mann algebra}

Having available the exterior differential calculus over a Gra{\ss}mann algebra, we can now turn to the discussion of its Connes' characters \cite{Conn94}. These are maps
$\int:\Omega^p(G^n)\to \mathds{R}$ subject to the conditions
\begin{equation}\label{bcch35b-d}
\int d\omega_{p-1}=0\qquad\qquad\qquad\text{and}\qquad\qquad\qquad
\int[\omega_r,\omega_{p-r}]=0\quad:\,p\geq r.
\end{equation}
Some authors also include $\int \omega_{p}=0$ for all $p<n$ 
amongst the definitions, but we do not. Since $\Omega^{\displaystyle\cdot}(G^n)$ is graded commutative, the second of the conditions \eqref{bcch35b-d} is void; the protection of the map $\int$ is not needed. As proven, $p$-dimensional characters are in bijective correspondence with normalized cyclic Hochschild $p$-cocycles, and so their classification can be reduced to the normalized cyclic cohomology $HC^{\displaystyle\cdot}(G^n)$; recall that we use the different notation $H^{\displaystyle\cdot}_\lambda(G^n)$ for the cyclic cohomology. We repeat the relevant definitions, particularized to the case at hand; the boundary operator 
(see \eqref{boundopvarphi}) acts on a cochain $\varphi_p\in C^p(G^n)$ as
\begin{equation}\label{bcch35e}
b\,\varphi(f_0,f_1,\ldots,f_{p})=
\end{equation}
\begin{equation}\nonumber
\sum_{i=0}^{p-1}(-1)^{i+\sum\limits_{j=0}^i|f_j|}\varphi(f_0,\ldots,f_if_{i+1},\ldots,f_{p})
-(-1)^{(1+|f_{p}|)((p-1)+\sum\limits_{j=0}^{p-1}|f_j|)}\varphi(f_{p}\,f_0,f_1,\ldots,f_{p-1})
\end{equation}
and the cyclicity operator (see \eqref{cyclopvarphi}) as
\begin{equation}\label{bcch35f}
\lambda\,\varphi(f_0,f_1,\ldots,f_{p})=(-1)^{(1+|f_{p}|)(p+\sum\limits_{j=0}^{p-1}|f_j|)}
\varphi(f_{p},f_0,f_1,\ldots,f_{p-1}).
\end{equation}
Finally, the space of cochains $H^p(G^n)$ is equipped with a $\mathds{Z}/2$-grading; the grading operator is
\begin{equation}\label{bcch35g}
\varepsilon\,\varphi(f_0,f_1,\ldots,f_{p})=(-1)^{p+\sum\limits_{i=0}^{p}|f_i|)}
\varphi(f_0,f_1,\ldots,f_{p})
\end{equation}
which anticommutes with both the boundary operator and the cyclicity operator.
We reiterate, for emphasis, our sign conventions in \eqref{bcch35e} and \eqref{bcch35f} differ from those commonly used in the literature. 

Let us go through some examples; for the time being, the normalization condition is dispensed. We begin with the case $p=0$ for arbitrary $n$; the $\mathds{R}$-linear maps 
$\varphi:G^n\to\mathds{R}$ are just the elements of the algebraic dual of $G^n$, which has the same dimension 
$2^n$ as $G^n$ itself. Introducing the notation $I=(i_1,\ldots,i_p)$ for the ordered $p$-tupel with $i_1 < \cdots < i_p$, the dual basis $\varepsilon_I$ is defined by $\varepsilon_I(\xi^{I^{\prime}})=\delta_I{}^{I^{\prime}}$; for the unit $1$ we set $I=0$ so that $\varepsilon_{0}(1)=1$, and zero otherwise. A particular basis element is given by  $\varepsilon_{1,\ldots,n}$ with the properties $\varepsilon_{1,\ldots,n}(\xi^1,\ldots,\xi^n)=1$
and $\varepsilon_{1,\ldots,n}(\xi^{i_1},\ldots,\xi^{i_p})=0$ for all $0\leq p<n$; it reproduces the Berezin integral $J(f)$ since
\begin{equation}\label{bcch35h}
\varepsilon_{1,\ldots,n}(f)=f_{1\cdots n}.
\end{equation}
As we know, amongst the linear maps $J:G^n\to\mathds{R}$ it is singled out by translational invariance; beyond this mere recognition, however, no new insight can be drawn from this observation.

Next, we investigate the case $n=1$ with $p=1$. As one easily verifies, there is only one single cyclic $1$-cocycle $\varphi=\varepsilon_{1,1}$; hence, it is automatically normalized. In terms of the associated character, we thus have
\begin{equation}\label{bcch35j}
\varphi(f,g)=\int\,f\,dg=f_1g_1
\end{equation}
where $f(\xi)=f_0+f_1\xi$ and $g(\xi)=g_0+g_1\xi$, and so the outcome
\begin{equation}\label{bcch35k}
\int\,d\xi=0\qquad\qquad\qquad\int\,\xi\,d\xi=0
\end{equation}
is identical with the Berezin rules; this observation has already been made by Kastler (see the introduction in \cite{Kast88}), but the case $n=1$ with $p=1$ is rather special, as will be seen below.

The next and last case we consider is $n=2$ with $p\leq 2$, from which the general strategy will then be read off. This specific example was also treated by Coquereaux $\&$ Ragoucy \cite{Coqu95} 
(cf. \cite{Coqu90,Coqu91}), but our intention is different. Apart from the fact that we view $G^n$ as an algebra over $\mathds{R}$, and not over $\mathds{C}$, these authors disregard the normalization condition. The really essential point, however, is that we do not use the standard Koszul-Milnor-Quillen sign rules; though one would suspect this to make no essential difference, it will turn out to be of crucial importance, and so we are obliged to give some details. The case $p=0$ has already been commented upon, and so we can turn to $p=1$; with the notation $f(\xi)=f_0+f_1\xi^1+f_2\xi^2+f_{12}\xi^{12}$, the cyclic cocycles are:
\begin{align}\begin{split}\label{bcch35l}
\chi_{1,1}  &=\varepsilon_{1,1}                     \\
\chi_{1,2} &=\varepsilon_{1,2}+\varepsilon_{2,1} \\
\chi_{2,2}  &=\varepsilon_{2,2}                     \\
\chi_{1,12} &=\varepsilon_{1,12}+\varepsilon_{12,1} \\
\chi_{2,12} &=\varepsilon_{2,12}+\varepsilon_{12,2}.
\end{split}\end{align}
Why we have chosen the symbol $\chi$ for these coycles will become apparent in the next paragraph. Thus, the normalized cyclic cohomology group is 
\begin{equation}\label{bcch35m}
HC^1(G^2)=\mathds{R}^3\oplus \mathds{R}^2
\end{equation}
where the first direct summand denotes the even and the second the odd cocycles; none of these cocycles can be written as  coboundaries. Proceeding to $p=2$, which we expect to be the case of interest for the purposes we have in mind, one finds the cyclic cocycles: 
\begin{align}\begin{split}\label{bcch35n}
\varphi_{0,0,0}  &=\varepsilon_{0,0,0}   \\
\varphi_{0,0,1}  &=\varepsilon_{0,0,1}-\varepsilon_{0,1,0}+\varepsilon_{1,0,0}   \\
\varphi_{0,0,2}  &=\varepsilon_{0,0,2}-\varepsilon_{0,2,0}+\varepsilon_{2,0,0}   \\
\varphi_{0,0,12} &=\varepsilon_{0,0,12}+\varepsilon_{0,12,0}+\varepsilon_{12,0,0} \\
\chi_{1,1,1}  &=\varepsilon_{1,1,1}                     \\
\chi_{2,2,2}  &=\varepsilon_{2,2,2}                     \\
\chi_{1,1,2} &=\varepsilon_{1,1,2}+\varepsilon_{1,2,1}+\varepsilon_{2,1,1} \\
\chi_{2,2,1} &=\varepsilon_{2,2,1}+\varepsilon_{2,1,2}+\varepsilon_{1,2,2} \\
\chi_{1,1,12}&=\varepsilon_{1,1,12}+\varepsilon_{1,12,1}+\varepsilon_{12,1,1}\\
\chi_{2,2,12}&=\varepsilon_{2,2,12}+\varepsilon_{2,12,2}+\varepsilon_{12,2,2}\\
\chi_{1,2,12}&=\varepsilon_{1,2,12}+\varepsilon_{1,12,2}+\varepsilon_{12,1,2}+
\varepsilon_{2,1,12}+\varepsilon_{2,12,1}+\varepsilon_{12,2,1}\,.
\end{split} 
\end{align}
In total, there are $11$ nontrivial cocycles, i.e. $Z^2_{\lambda}(G^2)=\mathds{R}^5\oplus\mathds{R}^6$; but among these there are some that may be written as boundaries, namely
\begin{alignat*}{5}
\varphi_{0,0,1}  &=&-b\,\varphi_{0,1} \\
\varphi_{0,0,2}  &=&-b\,\varphi_{0,2} \\
\varphi_{0,0,12} &=&-b\,\varphi_{0,12}
\end{alignat*}
where the $\varphi$s on the right are all elements of $C^1_{\lambda}(G^n)$. Thus the cyclic cohomology group $H^2_{\lambda}=Z^2_{\lambda}/B^2_{\lambda}$ is $H^2_{\lambda}(G^2)=\mathds{R}^4\oplus\mathds{R}^4$; note that $\varphi_{1,1,1}$ may not be written as a  boundary $b\,\varphi_{1,1}$ since a cyclic $\varphi_{1,1}\in C^1_{\lambda}(G^2)$ vanishes. Passing to normalized cyclic cocycles, also $\varphi_{0,0,0}$ is eliminated, and so we end up with
\begin{equation}\label{bcch35o}
HC^2(G^2)=\mathds{R}^3\oplus \mathds{R}^4.
\end{equation}
Observe, with our conventions all elements of $HC^p(G^2)$ with $p\leq 2$ are completely symmetric in their labels. Had we chosen instead the standard sign conventions \eqref{KMQsignruleb} and 
\eqref{KMQsignrulelambda}, as in \cite{Coqu95} and elsewhere, then e.g. in the last cocycle $\chi_{1,2,12}$ the contributions $\varepsilon_{1,12,2}$ and $\varepsilon_{2,12,1}$ would acquire a minus sign in front; this innocuous alteration has drastic consequences, as will be shown now.

What we claim is that only our sign rules, as opposed to those of Koszul-Milnor-Quillen, are compatible with the defining properties of a Connes' character; this claim at least holds in the present case of a graded commutative algebra. For the proof, we look at the last normalized cyclic cochain in 
\eqref{bcch35n}, which is of special relevance:
\begin{equation}\label{bcch35p}
\chi_{1,2,12}(f,g,h)=(f_{1}g_{2}h_{12}+f_{2}g_{1}h_{12})\pm(f_{1}g_{12}h_{2}+f_{2}g_{12}h_{1})+
(f_{12}g_{1}h_{2}+f_{12}g_{2}h_{1}).
\end{equation}
Here, the upper signs are obtained with our sign rules, and the lower ones are those of 
Coquereaux $\&$ Ragoucy \cite{Coqu95} and others. We evaluate this cocycle on the basis elements $\xi^1,\xi^2$ and $\xi^{12}=\xi^1\xi^2$ in two different orderings:
\begin{equation}\label{bcch35q}
\chi_{1,2,12}(\xi^{12},\xi^1,\xi^2)=1\qquad\qquad\qquad\chi_{1,2,12}(\xi^{1},\xi^{12},\xi^{2})=
\pm 1
\end{equation}
Now we invoke that $\chi_{1,2,12}$ may also be viewed as the character $\chi_{1,2,12}(f,g,h)=\int_{1,2,12} f\,dg\,dh$, and so we can continue with 
\begin{equation}\label{bcch35r}
\chi_{1,2,12}(\xi^{12},\xi^1,\xi^2)=1=\int_{1,2,12}\xi^{1}\xi^{2}d\xi^1d\xi^2
\end{equation}
in the first case, and in the second with
\begin{equation}\label{bcch35s}
\chi_{1,2,12}(\xi^{1},\xi^{12},\xi^{2})=
\pm 1=\int_{1,2,12}\xi^{1}d\xi^{12}d\xi^{2}=
\int_{1,2,12}\xi^{1}(d\xi^{1}\xi^2-\xi^{1}d\xi^2)d\xi^{2}=\int_{1,2,12}\xi^{1}\xi^2d\xi^{1}d\xi^{2}
\end{equation}
since for the last term in round brackets the integrand vanishes; whence, these two results are compatible for the upper sign only, as contended.

It will prove to be important for the further development that, as eq. \eqref{bcch35r} exhibits, the above character may also be interpreted as the Berezin integral for $n=2$; one thus expects that the Berezin integral calculus over $G^n$ will  intimately be related to the theory of Connes' characters over that algebra. Before we can turn to a closer analysis of this conjecture, however, we need a further concept to be introduced next.  

\section{Gra{\ss}mann currents}

We again assume, in an intermediate step, the (algebraic) Berezin rules to be given beforehand; the only ones  we accept are (see \eqref{bcch9b})
\begin{equation}\label{bcch36}
\int d^{\,n}\xi\,\xi^{i_1}\cdots\xi^{i_p}\,=
\begin{cases}
0&:\,p<n\\
\varepsilon^{i_1\cdots i_n}&:\,p=n
\end{cases}
\end{equation}
where the position of the `volume element' $d^{\,n}\xi$, which still awaits a proper definition, does not matter. Let us return to the pairing; hence, a metric is not needed. So recall, the dual $\Omega_p(G^n)$ of the space of $p$-forms 
$\Omega^p(G^n)$ is built from $p$-currents through (see \eqref{bcch22})
\begin{equation}\label{bcch37}
\langle\chi^p|\omega_p\rangle=\int\,d^{\,n}\xi\,
\frac{1}{p\,!}\chi^{i_1\cdots i_p}(\xi)^\ast\,
\omega_{i_1\cdots i_p}(\xi)
\end{equation}
where the components $\chi^{i_1\cdots i_p}$ of a $p$-current are completely symmetric in their indices.
In the classical situation, the dual of $\Omega^p(C^{\infty}(M))$ over a smooth manifold $M$ is the space of de Rham currents (see \cite{Dieu72}); in particular, for $p=0$ the dual includes the distributions. But for the case at hand such finesses are not needed. The adjoint $\bar{d}$ of the operator $d$, now being denoted as the divergence $\partial$, is   
\begin{equation}\label{bcch38-43}
(\partial\chi^p)^{i_2\cdots i_p}=\partial_{i_1}\chi^{i_1i_2\cdots i_p}.
\end{equation}
Accordingly, a current is said to be closed if $\partial\chi^p=0$, i.e. if its divergence is zero. We thus have available the dual $(\Omega^{}_{\displaystyle\cdot}(G^n),\partial)$ of the complex $(\Omega_{}^{\displaystyle\cdot}(G^n),d)$; since the latter is acyclic, also the complex of currents is.
This is a general property, namely, given a cochain complex $(C_{}^{\displaystyle\cdot},d)$ over the real or complex numbers which is acyclic, then also the dual chain complex $(C^{}_{\displaystyle\cdot},\bar{d})$ is acyclic, and conversely.

We expect the latter proposition to be known to the experts; however this may be, the proof is sufficiently simple to be given. So let $V$ be linear spaces over $\mathds{R}$ or $\mathds{C}$, and $V^{\,\prime}$ its dual. For $U\leq V$ any subspace of $V$, the orthogonal complement $U^{\perp}$ of $U$ in $V^{\,\prime}$ is defined by $U^{\perp}=\{\varphi\in V^{\,\prime}|\varphi(u)=0\,\,\mathrm{for\,\,all}\,\,u\in U\}$, and similarly for a subspace $U^{\,\prime}\leq V^{\,\prime}$; one also has $U^{\perp\perp}=U$. Furthermore,  if $\phi: V\to W$ is a linear mapping into a second vector space $W$ and $\phi^{\,\prime}: W^{\,\prime}\to V^{\,\prime}$ its dual, then the following identities  
$(\mathrm{Ker}\,\phi^{\,\prime})^{\perp}=\mathrm{Im}\,\phi$ and 
$(\mathrm{Ker}\,\phi)^{\perp}=\mathrm{Im}\,\phi^{\,\prime}$ hold.
Suppose now that the sequence of linear maps
$U\overset{\chi}{\longrightarrow}V\overset{\phi}{\longrightarrow}W$
is exact at $V$; then also the dual sequence
$U^{\,\prime}\underset{\chi^{\ast}}{\longleftarrow}V^{\,\prime}\underset{\phi^{\,\prime}}{\longleftarrow}
W^{\,\prime}$
is exact at $V^{\,\prime}$. In fact, since $\mathrm{Im}\,\chi=\mathrm{Ker}\,\phi$ we have
$\mathrm{Ker}\,\phi^{\,\prime}=(\mathrm{Im}\,\phi)^{\perp}=(\mathrm{Ker}\,\phi)^{\perp}=
\mathrm{Im}\,\phi^{\,\prime}$
as was to be proven.

From currents, one can construct associated Hochschild cochains in a rather obvious way 
(cf. also \cite{Coqu95}); we simply set
\begin{equation}\label{bcch44}
\chi(f_0,\ldots,f_p)=\chi(f_0\,df_1\vee\cdots\vee df_p)=
\int\,d^{\,n}\xi\,(\chi^{i_1\cdots i_p})^{\ast}f_0\,\partial_{i_1}f_1\cdots\partial_{i_p}f_p.
\end{equation}
Here the first equality sign is meant as a definition, and the second makes sense since the left hand side is graded symmetric in the $f_i$ with $i\in\{1,\ldots,n\}$, as is the right hand side because $\chi^{i_1\cdots i_p}$ is symmetric in the ordinary sense; whence, the graded symmetric tensor product defined in Subsec. \ref{Thegradedcommutativecase} enters  
in an essential way. 

According to the definition \eqref{defboundopgng}, the boundary of the Hochschild cochain \eqref{bcch44} 
is given by $b\,\chi(\omega_p)=\chi(b(\omega))$ where $(\chi|\omega)=\chi(\omega)$ is the evaluation map, and since $\Omega^pG^n$ is graded commutative, $\chi(f_0,\ldots,f_p)$ is a cocycle. Note that this conclusion is  rather immediate since our definition relies on the property \eqref{defboundopgng}, which is not shared by the alternative choice \eqref{KMQsignruleb} used in the literature. Hence, if a Hochschild cochain may be represented through a current, then this cochain is a cocycle. 

Let us restrict our considerations to closed currents in what follows; we then show that the associated cocycles are also cyclic. As for the proof, since $\Omega^p(G^n)$ algebra is graded commutative, we have
\begin{align*}
\chi(f_0\,df_1\vee\cdots\vee df_p)&=(-1)^{|df_p|(|f_0|+\sum\limits_{i=1}^{p-1}|df_i|)}\chi(df_p\vee f_0\,df_1
\vee\cdots\vee 
df_{p-1})\\
&=(-1)^{|df_p|(|f_0|+\sum\limits_{i=1}^{p-1}|df_i|)}\chi(d(f_pf_0)\vee df_1\vee\cdots 
\vee df_{p-1}-(-1)^{|f_p|}f_pdf_0\vee df_1\vee\cdots\vee df_{p-1})
\end{align*}
where, since $\chi^p$ is closed, the first term vanishes. We thus end up with
$$\chi(f_0\,df_1\vee\cdots\vee df_p)=(-1)^{|df_p|\sum\limits_{i=0}^{p-1}|df_i|}\chi(f_p\,df_0\vee\,df_1\vee
\cdots\vee df_{p-1})$$
and comparing this with the definition \eqref{cyclopkdefgncg} of the cyclicity operator $\lambda$, we conclude that $\lambda\chi^p=\chi^p$ as claimed.

Finally, the Hochschild cocycles associated to closed currents are also normalized. Indeed, the Berezin rules imply $\chi(1,f_1,\ldots,f_p)=0$, and cyclicity then entails the normalization property.

To summarize, via the map $\chi^p:\Omega^p(G^n)\to\mathds{R}$ a closed $p$-current determines a $p$-dimensional character over the Gra{\ss}mann algebra. 

What one would like to have then is, every character may be represented by a closed current. In the classical de Rham case over a compact manifold, Connes (\cite{Conn85}) has proven that this is indeed true. But in the graded commutative case, the corresponding proof does not go through (contrary to the claim made in 
\cite{Coqu95}) since in
\begin{equation*}
\chi(f_0\,df_1\vee\cdots\vee df_p)=
\int\,d^{\,n}\xi\,\,(\chi^{i_1\cdots i_p})^{\ast}\,\frac{1}{p\,!}\,\sum_{\pi\in S_p}f_0\,\partial_{i_{\pi(1)}}f_1\cdots
\partial_{i_{\pi(p)}}f_p
\end{equation*}
the symmetrization may not be shifted to the $f_i\,$s with $i\in\{1,\ldots,n\}$ because the latter only  commute in the graded sense.

In order to gain a feeling for what one may expect, let us return to the example with $n=2$ and $p\leq 2$ of the previous paragraph; the normalized cyclic cocycles that can not be written as coboundaries are denoted as $\chi$ in eqs. 
\eqref{bcch35l} and \eqref{bcch35n} there. The assertion then is, all these cocycles may be expressed in terms of closed currents; hence, at least in this particular case the above conjecture is indeed true. 

For the verification we begin with $p=0$; this is the dual $(G^2)^{\prime}$, were it not for the normalization condition. Here the request for closedness of the current is void, and with
\begin{equation}\label{bcch4}
\chi(f)=\int\,d^{\,2}\xi\,\chi(\xi)^{\ast}f(\xi)
\end{equation}
one finds, in an obvious notation:
\begin{alignat*}{5}
\chi_{1}(f)&=f_1\qquad\qquad&:\,&\chi_1(\xi)&=&-\xi^2 \\
\chi_{2}(f)&=f_2\qquad\qquad&:\,&\chi_2(\xi)&=&+\xi^1 \\
\chi_{12}(f)&=f_{12}\qquad\qquad&:\,&\chi_{12}(\xi)&=&+1. 
\end{alignat*}
The result for $p=1$ is more interesting; in the notation of \eqref{bcch44} and with $(\chi^i)_{i=1,2}=\chi$ one has:
\begin{alignat*}{5}
\chi_{1,1}(f,g)&=f_1g_1\qquad\qquad&:\,&\chi_{1,1}(\xi)&=&
\begin{pmatrix}-\xi^2\\0\end{pmatrix}\\ 
\chi_{2,2}(f,g)&=f_2g_2\qquad\qquad&:\,&\chi_{2,2}(\xi)&=&
\begin{pmatrix}0\\+\xi^1\end{pmatrix} \\
\chi_{1,2}(f,g)&=f_1g_2+f_2g_1\qquad\qquad&:\,&\chi_{1,2}(\xi)&=
&\begin{pmatrix}+\xi^1\\-\xi^2\end{pmatrix}\\
\chi_{1,12}(f,g)&=f_{1}g_{12}+f_{12}g_{1}\qquad\qquad&:\,&\chi_{1,12}(\xi)&=&
\begin{pmatrix}1\\0\end{pmatrix} \\
\chi_{2,12}(f,g)&=f_{2}g_{12}+f_{12}g_{2}\qquad\qquad&:\,&\chi_{2,12}(\xi)&=&
\begin{pmatrix}0\\1\end{pmatrix} 
\end{alignat*}
and one checks that all these currents are closed. Finally, for $p=2$ and in matrix notation $(\chi^{ij})_{i,j=1,2}=\chi$ we obtain:
\begin{alignat*}{5}
\chi_{1,1,1}(f,g,h)&=f_1g_1h_1\qquad&:\,&\chi_{1,1,1}(\xi)&=&
\begin{pmatrix}-\xi^2&0\\0&0\end{pmatrix}\\ 
\chi_{2,2,2}(f,g,h)&=f_2g_2h_2\qquad&:\,&\chi_{2,2,2}(\xi)&=&
\begin{pmatrix}0&0\\0&+\xi^1\end{pmatrix} \\
\chi_{1,1,2}(f,g,h)&=f_1g_1h_2+f_1g_2h_1+f_2g_1h_1\qquad&:\,&\chi_{1,1,2}(\xi)&=
&\begin{pmatrix}+\xi^1&-\xi^2\\-\xi^2&0\end{pmatrix}\\
\chi_{2,2,1}(f,g,h)&=f_2g_2h_1+f_2g_1h_2+f_1g_2h_2\qquad&:\,&\chi_{2,2,1}(\xi)&=
&\begin{pmatrix}0&+\xi^1\\+\xi^1&-\xi^2\end{pmatrix}\\
\chi_{1,1,12}(f,g,h)&=f_{1}g_{1}h_{12}+f_{1}g_{12}h_{1}+f_{12}g_{1}h_{1}\qquad&:
\,&\chi_{1,1,12}(\xi)&=&
\begin{pmatrix}1&0\\0&0\end{pmatrix} \\
\chi_{2,2,12}(f,g,h)&=f_{2}g_{2}h_{12}+f_{2}g_{12}h_{2}+f_{12}g_{2}h_{2}
\qquad&:\,&\chi_{2,2,12}(\xi)&=&
\begin{pmatrix}0&0\\0&1\end{pmatrix} \\
\chi_{1,2,12}(f,g,h)&=f_{1}g_{2}h_{12}+f_{2}g_{1}h_{12}+f_{1}g_{12}h_{2}+f_{2}g_{12}h_{1}+
f_{12}g_{1}h_{2}+f_{12}g_{2}h_{1}\qquad&:\,&\chi_{1,2,12}(\xi)&=&
\begin{pmatrix}0&1\\1&0\end{pmatrix}. 
\end{alignat*}
Again, it is rather obvious that all these currents are closed. The last cocycle, denoted $\tau_{\text{vol}}=\chi_{1,2,12}$, will be the one of ultimate interest, and we restate it here in the form of a Berezin integral 
\begin{equation}\label{bcch45}
\tau_{\text{vol}}(f,g,h)=
\int\,d^{\,2}\xi\,(f\,\partial_{1}g\,\partial_{2}h+f\,\partial_{2}g\,\partial_{1}h)
\end{equation}
for later reference.

Let us remark, these latter results could not have been found by Coquereaux \& Ragoucy \cite{Coqu95} since their sign rules are not compatible with the Berezin rules.

From what we have learned for $n=2$, one expects the general situation to be that all nontrivial characters may be represented in terms of currents; indeed, this expectation is correct and its verification is given below. But we have no direct proof for this claim; instead, we proceed indirectly and invoke that the dimensions of the cohomology groups $H^p_{\lambda}(G^n)$  are known. They were derived by Kassel \cite{Kass86} by means of a K\"unneth type argument and given in the form of a $\mathds{Z}/2$-graded 
Poincar\'e polynomial; but we do not state his result here since it will emerge in the course of the proof. The approach given below relies in part on ideas developed in 
\cite{Coqu95}; they require revision, however, and also our conclusions are different.

Up to now we only know that a closed current $\chi^p\in\mathrm{Ker}\,\partial_p$ determines a normalized element of $Z^p_{\lambda}(G^n)$, but we do not have under control whether there are elements of $B^p_{\lambda}(G^n)$ among these. By a counting argument we shall show that the dimension of the vector space  $\mathrm{Ker}\,\partial_p$ is identical with the dimension of the space $HC^p(G^n)$; hence, these spaces are isomorphic and thus the map
$\mathrm{Ker}\,\partial_p\to HC^p$ defined by \eqref{bcch44}
is, in particular, a surjection. This final result may be paraphrased by saying that those normalized cyclic cocycles that can be represented by a nonzero current may not be written as boundaries; hence, such currents compute the normalized cyclic cohomology. 

As to the proof, the strategy will be to establish a recurrence relation for the dimension of the space of closed $p$-currents. We begin with the dimension of the space $\mathcal{C}_p\doteq\Omega_p(G^n)$ of currents, which is
\begin{equation}\label{bcch45a}
\text{dim}\,\mathcal{C}_p=2^n{n+p-1 \choose p}\qquad\qquad\qquad:\,p\geq 0.
\end{equation}
For $p>1$, the vector space $\mathcal{C}_p$ can be decomposed into the direct sum $\mathcal{C}_p=\mathcal{C}\mathcal{C}_p\oplus\not{\!\mathcal{C}}\mathcal{C}_p$ of closed currents, i.e. $\mathcal{C}\mathcal{C}_p\doteq\mathrm{Ker}\,\partial_p$, and those which are not, i.e.  $\not{\!\mathcal{C}}\mathcal{C}_p\doteq\mathrm{Im}\,\partial_p$. Because $\partial_{p-1}\partial_p=0$, we thus have the inclusion $\not{\!\mathcal{C}}\mathcal{C}_p\subseteq\mathcal{C}\mathcal{C}_{p-1}$; as we have seen above, however, this is even an equality since together with the complex $\Omega^{\displaystyle\cdot}G^n$ also the dual complex of currents $\Omega_{\displaystyle\cdot}G^n$ is acyclic, and so
\begin{equation}\label{bcch45b}
\mathcal{C}_p=\mathcal{C}\mathcal{C}_p\oplus \mathcal{C}\mathcal{C}_{p-1}\qquad\qquad\qquad:\,p>1
\end{equation}
giving the recursion $\text{dim}(\mathcal{C}\mathcal{C}_p)=
\text{dim}(\mathcal{C}_p)-\text{dim}(\mathcal{C}\mathcal{C}_{p-1})$. For $p=1$, the complementary subspace of $\mathcal{C}\mathcal{C}_1$ in $\mathcal{C}_1$ consists of those elements of $G^n$, in which the top component is absent; using the notation $\mathcal{C}\mathcal{C}_{0}\doteq\not{\!\mathcal{C}}\mathcal{C}_1$, we thus have the decomposition 
$\mathcal{C}_1=\mathcal{C}\mathcal{C}_1\oplus\,\mathcal{C}\mathcal{C}_0$, where  $\text{dim}(\mathcal{C}\mathcal{C}_0)=2^n-1$. In this way the recursion is valid down to $p=1$, and  iteration yields 
\begin{equation}\label{bcch45c}
\text{dim}(\mathcal{C}\mathcal{C}_p)=\sum_{i=0}^p(-1)^i\,\text{dim}(\mathcal{C}_{p-i})-(-1)^p.
\end{equation}
The sum can be manipulated by means of the expansion
\begin{equation*}
\frac{1}{(1+t)(1-t)^n}=\sum_{p=0}^{\infty}\,t^p
\sum_{i=0}^{p}(-1)^i{n+p-i-1\choose p-i}\qquad\qquad\qquad:\,|t|<1.
\end{equation*}
Whence, on using that the elements of $\mathcal{C}\mathcal{C}_p$ divide into even and odd ones, the generating function for the dimensions of the closed currents $\mathcal{C}\mathcal{C}_p$ is the Poincar\'e polynomial
\begin{equation}\label{bcch45f}
P^n(t,\theta)=\frac{2^{n-1}(1+\theta)}{(1+t)(1-t)^n}-\frac{1}{1+t}
\end{equation}
where $\theta$ denotes the nontrivial generator of $\mathds{Z}/2$. In particular, for $n=2$ this  confirms the values determined by direct computation in the previous paragraph. 

We can now address the discussion of the results for the cyclic cohomology $H^{\displaystyle\cdot}_{\lambda}(G^n)$, as obtained by Kassel in \cite{Kass86}. In that work the cyclic cohomology group is written as the direct sum of two terms, the first of which is just specified by the Poincar\'e polynomial \eqref{bcch45f}; hence, according to what we have shown, this part can thus be characterized as the normalized cyclic cohomology 
$HC^{\displaystyle\cdot}(G^n)$. The second direct summand is the cyclic cohomology group of the real 
numbers $H^{\displaystyle\cdot}_{\lambda}(\mathds{R})$, which is trivial for $p$ odd and equal to 
$\mathds{R}$ for $p$ even (see, e.g. \cite{Rose94}), and so the result given in \cite{Kass86} can be read as
\begin{equation}\label{bcch45g}
H^{\displaystyle\cdot}_{\lambda}(G^n)=
HC^{\displaystyle\cdot}(G^n)\oplus H^{\displaystyle\cdot}_{\lambda}(\mathds{R}).
\end{equation}
What remains is to characterize the second direct summand as (non normalized) nontrivial cyclic 
cocycles. This latter part we identify with the cocycles
$$\varphi_{0\cdots0}(f_0,\ldots,f_p)=\varepsilon_{0\cdots0}(f_0,\ldots,f_p)$$
which, on imposing the cyclicity property, are nonvanishing only for $p$ even; of course, they can not be 
written in the form of a Berezin integral.

To summarize, in one aspect the above results go beyond those obtained in \cite{Kass86} since the cocycles belonging to the two direct summands can even be given in explicit form. 

\section{Berezin's integral as a Connes' character}

With this interlude on Gra{\ss}mann currents behind us, we now forget again about Berezin integration since a 
precise definition of the symbol $d^{\,n}\xi$ is not yet available. So let us return to characters on 
$G^n$; in this setting we can give a definite meaning to that symbol, and what we want is to relate 
somehow such Connes' characters to a Berezin type of integral. 

Recall, for $n=2$ and $p=2$ there are seven 
nontrivial characters and, as we have seen, this number grows rapidly with $n$; what is more, there are 
also characters for $p>n$, contrary to the classical case. Amongst this multitude of available characters, 
we claim that there is only one which  deserves to be named a `volume integral'. 

Such a construct is uniquely singled out by three natural defining properties. The first consists in the 
requirement that only $n$-dimensional characters 
\begin{equation}\label{bcch46}
\int:\Omega^n(G^n)\to \mathds{R}
\end{equation} 
are to be considered. At second, it should not make sense that a variable is integrated twice. Hence, a
`volume character' $\int$ 
is assumed to take nonzero values only on that subspace of $n$-forms, for which on the right hand side of
\begin{equation}\label{bcch47}
\int \omega_n=\int \frac{1}{n\,!}\,\omega_{i_1\cdots i_n}\,d\xi^{i_1}\cdots d\xi^{i_n}
\end{equation}
all the differentials are different, i.e. if $(i_1,\ldots,i_n)$ is a permutation of $(1,\ldots,n)$; 
whence, formally the axiom is
\begin{equation}\label{bcch48}
\int\omega_{n}=\int\frac{1}{n\,!}\,\sum_{\sigma\in S_n} \omega_{\sigma(1)\cdots\sigma(n)}
\,d\xi^{1}\cdots d\xi^{n}=\int\,\omega_{1\cdots n}\,d\xi^{1}\cdots d\xi^{n}
\end{equation}
since all other contributions vanish by definition, and since the differentials commute. 
Note that it would not suffice here to restrict the $\Omega^p(G^n)$ themselves to the corresponding subspaces since they are not invariant against exterior differentiation; instead, one must restrict the admissible maps $\int$, i.e. the characters.
For $\omega_n$ of the form $f_0\,df_1\cdots df_n$, we thus arrive at 
\begin{equation}\label{bcch49}
\tau_{\text{vol}}(f_0,\ldots,f_n)=\int\sum_{\sigma\in S_n} f_0\partial_{\sigma(1)}f_1
\cdots\partial_{\sigma(n)}f_n\,d\xi^{1}\cdots d\xi^{n}
\end{equation}
which gives us a special Hochschild $n$-cochain in terms of that $n$-dimensional character. 

The  definition, however,  is still not complete;
what finally remains is to assign the right hand side of \eqref{bcch49} a real number, for arbitrary 
$f$s in $G^n$. We do this by the prescription that it be evaluated by means of the 
formal Berezin integration rules, which is the third requirement.

The claim then is, the construct (cf. \eqref{bcch45} for $n=2$) indeed defines a Connes character. This  assertion, if valid, implies 
that the Berezin integration rules for Gra{\ss}mann variables and the defining properties of this special $n$-dimensional Connes' character 
do indeed match. After all, at the risk of being pedantic, the symbol $d^{\,n}\xi$ can then legitimately be viewed as the product $d\xi^1\cdots d\xi^n$ of the differentials; this does not mean, however, that the integral could be defined by iteration, as often claimed in the literature.

For the verification we must show, the cochain \eqref{bcch49} is a normalized cyclic cocycle. This is 
done on rewriting $\tau_{\text{vol}}$ in the form
\begin{equation}\label{bcch50}
\tau_{\text{vol}}(f_0,\ldots,f_n)=\int\,d^{\,n}\xi\,\chi^{i_1\cdots i_n}f_0 
\partial_{i_1}f_{1}\cdots\partial_{i_n}f_n
\end{equation}
where
\begin{equation}\label{bcch51}
\chi^{i_1\cdots i_n}=\sum_{\sigma\in S_n}\delta^{i_1}{}_{\sigma(1)}\cdots\delta^{i_n}{}_{\sigma(n)}
\end{equation}
is recognized as a constant, whence closed current. Now we are back at the situation of the previous 
section so that we can resort to the results obtained there; hence, we  are entitled to identify 
$\tau_{\text{vol}}$ as a normalized cyclic cocycle, i.e. a character. Furthermore, the corresponding 
normalized cyclic cocycle is not trivial, i.e. not a coboundary; of course, this fact is crucial. Let us 
also note, the property $\int d\omega_{n-1}=0$ of a character implies $\int d^{\,n}\xi\,\partial_if(\xi)=0$ 
for all $i\in\{1,\ldots,n\}$, which is one of the Berezin rules. The definition \eqref{bcch49} may be viewed as a generalization of the Berezin integral calculus on the Gra{\ss}mann algebra, relating it to the Connes' characters of noncommutative geometry.

Hence, eq. \eqref{bcch49} defines a special $n$-dimensional character if the right hand side is 
understood to be evaluated by means of the Berezin rules. The conventional Berezin integral over an 
element $f\in G^n$ is regained for $f_0=f$ and $f_i=\xi^i$ with
$i\in(1,\ldots,n)$, viz.
\begin{equation}\label{bcch52}
\tau_{\text{vol}}(f,\xi^1,\ldots,\xi^n)=\int\,d^{\,n}\xi\,f(\xi)=f_{1\cdots n}.
\end{equation}
After all, this is the final justification of the formula \eqref{bcch9c} we began with, i.e. $J(f)=\int d^{\,n}\xi\,f(\xi)$, and so the circle of ideas closes here, at the very end. 

To resume, Berezin integration over a Gra{\ss}mann algebra fits perfectly into the scheme of noncommutative geometry in that amongst Connes' integral calculi one unique normalized cyclic Hochschild cocycle can naturally be singled out, which reduces to the Berezin integral and is compatible with its defining rules. 

\section{Conclusion}

Beyond the Berezin rules \eqref{bcch9b} or \eqref{bcch9b'}, that we consider as safe, there is a variety of other rules around in the literature. One such example, given in Berezin's book \cite{Bere66} and elsewhere, is
$$\int d\xi^i\,\xi^j\overset{?}{=}\delta^{ij}$$
which we refuse since it would entail that the integral over anticommuting variables could be defined iteratively. Another one, also often being taken for granted, is 
$$d\xi^i\,d\xi^j+d\xi^j\,d\xi^i\overset{?}{=}0$$
so that the differentials anticommute; but this condition is at variance with the demand that the exterior differentiation $d$ be of grade $+1$, i.e. be an odd operator. Anticommuting differentials are suggested also by another rule frequently being stated, namely, that the `volume element' should transform under a linear transformation $\xi \mapsto \xi'=A\xi$ with $A\in\text{GL}(n,\mathds{R})$ as 
$$d^{\,n}\xi'\overset{?}{=}\text{det}(A)^{-1}d^{\,n}\xi.$$
This requirement is inspired by the rules \eqref{bcch9b'} and \eqref{bcch9b''} derived above - as far as one is willing at all to ascribe a transformation law to the volume element. Indeed, it requires anticommuting differentials; but instead of the determinant, its inverse appears in this latter formula, The way out being favoured in the literature is to define integration in the Gra{\ss}mann case as an operation over the tangent 
space, instead of the cotangent space in the conventional case, and so one invents another volume element
(see \cite{Bruz87}, and literature cited there)
$$d^{\,n}{}_{\!\!\textstyle\xi}\overset{?}{=}\frac{\partial}{\partial\xi^1}\wedge\cdots\wedge
\frac{\partial}{\partial\xi^n}$$
where the ungraded wedge product should be understood; it indeed reproduces the postulated transformation law. According to our conviction, however, deep water is reached at this point. As we believe, the idea to ascribe a transformation law to
the volume element must be relinquished; actually, there is also no need to do so since Berezin integration acts as a differential operator, and the inverse of the determinant merely follows from the fact that the partial differentiations anticommute. Thus, the validity of eq. \eqref{bcch9b''} is guaranteed without further ado, and this is the only property that enters the fermionic path integral in the representation through coherent states; the transition to the complex case poses no added difficulty. Also, in the supersymmetric situation, the property \eqref{bcch9b''} is sufficient to define an integral of the form 
$$\int d^{\,m}x\int d^{\,n}\xi\,\, f(x,\xi)=\int d^{\,m}x\,\frac{\partial}{\partial\xi^n\cdots\partial\xi^1}\, f(x,\xi)$$
over the superspace $\mathds{R}^{m,n}$; under a change of supercoordinates, the differentiations then provide for the correct transformation law so that the superdeterminant of the Jacobian is reproduced correctly. A similar point of view is advocated by Cartier et al. \cite{Cart02}, being almost in line with our arguments.  In this way, the above apparent discrepancy is resolved, without any need to alter or supplement the rules taken for granted in the main text.
\newpage
\pagenumbering{Roman}\setcounter{page}{1}


\begin{thebibliography}{99}

\footnotesize
\bibitem{Itzy80}
Itzykson C. and Zuber J. B., \emph{Quantum Field Theory}, McGraw-Hill, New York 1980.
\bibitem{Ryde85}
Ryder L. H., \emph{Quantum Field Theory}, Cambr. Univ. Press, Cambridge 1985.
\bibitem{Nieu81}
Nieuwenhuizen, P. van, "Supergravity", Phys. Rep. \textbf{68} (1981) 189-398.
\bibitem{Wess83}
Wess J. and Bagger J., \emph{Supersymmetry and Supergravity}, Princeton Univ. Press, Princeton 1983.
\bibitem{Schw53}
Schwinger J., "A note on the quantum dynamical principle", Phil. Mag. \textbf{44} (1953) 1171-1179.
\bibitem{Schw70}
Schwinger J., \emph{Quantum Kinematics and Dynamics}, Benjamin, New York 1970.
\bibitem{Matt55}
Matthews P. T. and Salam A., "Propagators of quantized fields", Nuovo Cim. \textbf{2} (1955) 120-134.
\bibitem{Bere66}
F. A. Berezin, \textit{The Method of Second Quantization}, Academic Press, New York 1966. 
\bibitem{Bere79}
Berezin F. A., "Differential forms on supermanifolds", Sov. J. Nucl. Phys. \textbf{30} (1979) 605-609.
\bibitem{Bere87}
Berezin F. A., \emph{Introduction to Superanalysis}, Reidel, Dordrecht 1987.
\bibitem{Conn94}
Connes A., \emph{Noncommutative Geometry}, Academic Press, London and San Diego
1994.
\bibitem{Kast88}
Kastler D., \emph{Cyclic Cohomology within the Differential Envelope}, Hermann, Paris 1988.
\bibitem{Karo82}
Karoubi M., "Connexions, courbures et classes charact\'eristiques en K-th\'eorie alg\'ebrique", Canadian Math. Soc. Proc. \textbf{2} (1982) 19-27.
\bibitem{Coqu90}
Coquereaux R., Frappat L., Ragoucy E. and Sorba P., "Extended super-Ka\v{c}-Moody algebras and their super-derivation algebras", Comm. Math. Phys. \textbf{133} (1990) 1-35.
\bibitem{Coqu91}
Coquereaux R., Jadczyk A. and Kastler D., "Differential and integral calculus of Grassmann algebras", Rev. Math. Phys. \textbf{3} (1991) 63-99.
\bibitem{Coqu95}
Coquereaux R. and Ragoucy E., "Currents on Grassmann algebras", J. Geom. Phys. 
\textbf{15} (1995) 333-352.
\bibitem{Conn85}
Connes A., "Non-commutative differential geometry", Publ. Math. IHES \textbf{62} (1985) 41-144.
\bibitem{Land97}
Landi G., \emph{An Introduction to Noncommutative Spaces and their Geometries},
Lecture Notes in Physics
$\bf{51}$, Springer, Berlin 1997.
\bibitem{Grac01}
Gracia-Bondia J. M., V\'{a}rilly J. C. and Figueroa H., \emph{Elements of
Noncommutative Geometry},
Birkh\"auser, Boston 2001.
\bibitem{Kass86}
Kassel C., "A K\"unneth like formula for the cyclic cohomology of $\mathds{Z}/2$-graded algebras", Math. Ann. \textbf{275} (1986) 683-699.
\bibitem{Kast90}
Kastler D., "Introduction to entire cyclic cohomology (of $\mathds{Z}/2$-graded Banach algebras)", in \emph{Stochastics, Algebra and Analysis in Classical and Quantum Dynamics}, S. Albeverio, P. Blanchard and T. Testard Eds., Kluver Academic Publishers, Dordrecht 1990.
\bibitem{Brod98}
Brodzki J., \emph{An Introduction to K-theory and Cyclic Cohomology}, Polish Scientific Publishers (PWN), Warszawa 1998.
\bibitem{Loda92}
Loday J. L., \emph{Cyclic Homology}, Springer, Berlin 1992.
\bibitem{DeWi84}
DeWitt B. S., \textit{Supermanifolds}, Cambridge Univ. Press, Cambridge 1984.
\bibitem{Leit80}
Leites D. A., "Introduction to the theory of supermanifolds", Russ. Math. Surv. \textbf{35} (1980) 1-64.
\bibitem{Vlad84}
Vladimorov V. S. and Volovich I. V., "Superanalysis I: Differential calculus", 
Theor. Math. Phys. \textbf{59} (1984) 317-335;  
"Superanalysis II: Integral calculus", Theor. Math. Phys. \textbf{60} (1985) 743-765.
\bibitem{Cons94}
Constantinescu F. and de Groote H. F., \emph{Geometrische und algebraische Methoden in der Physik: Supermannigfaltigkeiten und Virasoro Algebren}, Teubner, Stuttgart 1994.
\bibitem{Deli99}
Deligne P., Etingof P., Freed D. S., Jeffrey L. C., Kazhdan D., Morgan J. W., Morrison D. R. and Witten E.  eds., \emph{Quantum Fields and Strings: A Course for Mathematicians}, Vol. I, Am. Math. Soc., Providence  
1999.
\bibitem{Guil99}
Guillemin V. W. and Sternberg S., \emph{Supersymmetry and Equivariant de Rham Theory}, Springer, Berlin 1999.
\bibitem{Bogn74}
Bognar J., \emph{Indefinite Inner Product Spaces}, Springer, Berlin 1974.
\bibitem{Gren02}
Grensing G., "On ghost fermions", Eur. Phys. J. C \textbf{23} (2002) 377-387.
\bibitem{Henn92}
Henneaux M. and Teitelboim C., \textit{Quantization of Gauge Systems}, Princeton Univ. 
Press, Princeton 1992. 
\bibitem{Bott82}
Bott R. and Tu L. W., \emph{Differential Forms in Algebraic Topology}, Springer, New York 1982.
\bibitem{Tayl96}
Taylor M. E., \emph{Partial Differential Equations} Vol. I, Springer, New York 1996.
\bibitem{Dieu72}
Dieudonn\'e J., \emph{Treatise on Analysis} Vol. III, Academic Press, New York 1972.
\bibitem{Rose94}
Rosenfeld J., \emph{Algebraic K-Theory and Its Applications}, Springer, New York 1994.
\bibitem{Bruz87}
Bruzzo U., "Integration on Supermanifolds", in \emph{General Relativity and Gravitational Physics}, U. Bruzzo, R. Cianci and E. Massa Eds., World Scientific, Singapore 1987.
\bibitem{Cart02}
Cartier P., DeWitt-Morette C., Ihl M. and S\"amann C., "Supermanifolds - application to supersymmetry", 
math-phys/0202026.

\end{thebibliography}
\end{document}